\documentclass[fleqn,usenatbib]{mnras}

\usepackage[T1]{fontenc}

\DeclareRobustCommand{\VAN}[3]{#2}
\let\VANthebibliography\thebibliography
\def\thebibliography{\DeclareRobustCommand{\VAN}[3]{##3}\VANthebibliography}

\usepackage{graphicx}	
\usepackage{amsmath}	
\usepackage{caption} 
\usepackage{color,soul}
\usepackage{lineno}

\usepackage{newtxtext,newtxmath}

\title[X-rays and $\gamma$-rays from globular clusters]{How the dynamical properties of globular clusters impact their $\gamma$-ray and X-ray emission}

\author[R. de Menezes et al.]{
Raniere de Menezes$^{1,2,3}$\thanks{E-mail: ranieremaciel.demenezes@unito.it}, Federico Di Pierro$^{1}$, and Andrea Chiavassa$^{1,2}$
\\
$^{1}$Istituto Nazionale di Fisica Nucleare, Sezione di Torino, Via Pietro Giuria 1, 10125 Turin, Italy\\
$^{2}$Dipartimento di Fisica, Universit\`a degli Studi di Torino, Via Pietro Giuria 1, 10125 Turin, Italy\\
$^{3}$Universidade de S\~ao Paulo, Departamento de Astronomia, Rua do Mat\~ao, 1226, S\~ao Paulo, SP 05508-090, Brazil\\
}

\date{Accepted 2023 June 05. Received 2023 June 02; in original form 2023 April 04}

\pubyear{2023}

\begin{document}
\label{firstpage}
\pagerange{\pageref{firstpage}--\pageref{lastpage}}
\maketitle

\begin{abstract}
The X-ray and $\gamma$-ray emission of globular clusters (GCs) is attributed to their large fraction of compact binary systems, especially those with millisecond pulsars (MSPs). We analyze a population of 124 Galactic GCs to investigate how their dynamical properties affect the formation and evolution of compact binary systems and how this can be translated into the clusters' observed X-ray and $\gamma$-ray emission. We use mainly \textit{Chandra} X-ray Observatory and \textit{Fermi} Large Area Telescope observations to achieve our goals and start by detecting 39 GCs in $\gamma$ rays, seven of which are not listed in previous \textit{Fermi}-LAT catalogs. Additionally, we find that the total number of X-ray sources within a GC and its $\gamma$-ray luminosity are linearly correlated with the stellar encounter rate, indicating that compact binary systems are mainly formed via close stellar encounters. We also find an unexpected rise in the number of X-ray sources for GCs with low rates of stellar encounters, suggesting that there is a dynamical threshold where the formation of X-ray sources is dominated by stellar encounters. Furthermore, we use the Heggie-Hills law to find that subsequent stellar encounters in these compact binaries will, on average, make the binaries even harder, with basically no possibility of binary ionization. Finally, we find that all GCs are point-like sources in $\gamma$ rays, indicating that the MSPs are concentrated in the clusters' cores, likely due to dynamical friction.
\end{abstract}



\section{Introduction}
\label{sec:intro}

Globular clusters (GCs) are evolved stellar systems typically with an exceptionally high number density of stars in their cores, often reaching more than 1000 pc$^{-3}$ \citep{Sollima2017global_mass_func}. This scenario makes dynamical interactions between stars a relatively common phenomenon, which can in turn lead to the formation (and possibly ionization) of compact binary systems \citep{pooley2003dynamical, verbunt2003binary}, necessary for the development of low-mass X-ray binaries (LMXB) that can eventually evolve into $\gamma$-ray emitting millisecond pulsars (MSPs). The formation of compact binaries in GCs is then expected to be related to the stellar encounter rate \citep{bahramian2013stellar}:

\begin{equation}
    \Gamma \propto \int{(n^2/\sigma)4\pi r^2dr},
    \label{eq:enc_rate}
\end{equation}
where $n = n(r)$ is the stellar number density, $\sigma = \sigma (r)$ is the stellar velocity dispersion, and the integral is performed over the volume of the cluster. The formation of tidally captured binaries in stellar encounters described by Eq. \ref{eq:enc_rate} is possible only when two stars pass so close to each other that the induced tides are able to dissipate the excess energy of the unbound orbital motion \citep{mardling1996tidal,mardling2001tidal}. For an encounter between a neutron star (with mass $m_{\mathrm{c}} \approx 1.5$ M$_{\odot}$) and a typical main sequence star in the core of a GC (with mass $m_{\star} \approx 0.5$ M$_{\odot}$ and radius R$_{\star}$) with central velocity dispersion in the range $\sigma_{\mathrm{c}} = 5 \sim 10$ km s$^{-1}$, the maximum periastron separation necessary for tidal capture must be around 3 -- 4.5 R$_{\star}$ \citep{mardling2001tidal}, depending also on the internal structure of both stars. The final orbit will be compact and highly eccentric, although the tides will tend to circularize it on a time scale that depends on the orbital semi-major axis (i.e. $t_{\mathrm{circ}} \propto a^8$) and on the internal structure of both stars \citep[for more details, see Eq. 4.13 in][]{zahn1977tidal}. For the binary configuration discussed above, if the original eccentricity of the system is in the range 0.5 -- 0.9, the circularization time scale will be $t_{\mathrm{circ}} \approx 10^6 \sim 10^9$ years, although it can be even smaller if one of the stars is a giant or subgiant \citep{verbunt1995tidal}.

Besides close stellar encounters, the formation of LMXBs (and eventually MSPs) in GCs requires that these clusters are able to retain a significant fraction of their neutron stars, which is achievable if these neutron stars are formed mainly via electron-capture supernovae \citep{nomoto1984evolution,nomoto1987evolution}, where the expected natal kicks are smaller than in core-collapse supernova, making it harder for the neutron stars to reach the escape velocity of the clusters \citep{Podsiadlowski2004effects,ivanova2008formation,claire2019millisecond}.

There are 37 GCs listed in the latest release of the \textit{Fermi} Large Area Telescope (LAT) source catalog \citep[4FGL-DR3;][]{abdollahi2020_4FGL,abdollahi20224fgl_incremental} and their $\gamma$-ray emission is mainly interpreted as coming from the populations of MSPs they host \citep{abdo2009detection47Tuc,venter2009predictions,abdo2010population,deMenezes2019milky}. These $\gamma$ rays are likely generated via curvature radiation in the outer magnetospheres of the MSPs, near the light cylinder \citep{harding2005high,caraveo2014gamma,kalapotharakos2019fundamental,Kalapotharakos2022Fundamental}. On the X-ray side, high-resolution observations with the \textit{Chandra} X-ray Observatory have revealed numerous sources in 72 GCs \citep{evans2010_CSC,evans2020_CSC2}, which are divided into many different classes, such as LMXBs, cataclysmic variables, MSPs, and coronally active main-sequence binaries.

The number of $\gamma$-ray and X-ray sources in GCs can, however, be affected by other dynamical factors beyond $\Gamma$. For instance, once a compact binary is formed, it may undergo subsequent encounters, which may i) ionize/excite the system, ii) shrink it even more, or even iii) exchange binary members. The rate at which a binary system undergoes close encounters with single stars is given by \citep{verbunt2003binary}:

\begin{equation}
    \Lambda \propto \frac{n}{\sigma}a,
    \label{eq:lambda}
\end{equation}
where $n$ is the local density of single stars, $\sigma$ is the local velocity dispersion, and $a$ is the semi-major orbital axis of the binary. The fate of the binary under subsequent encounters with typical stars in a GC (i.e., masses $m_{\star} \approx 0.5$ M$_{\odot}$) will be dictated by the Heggie-Hills law \citep{heggie1975binary,hills1975encounters}: hard binaries tend to get harder and soft binaries tend to get softer in face of three-body encounters. A hard binary in this context means a system with binding energy ($\epsilon$) higher than the average kinetic energy of the stars in the cluster, i.e. $|\epsilon| > m_{\star}\sigma^2$. If instead $|\epsilon| < m_{\star}\sigma^2$, we have a soft binary. On the other hand, exchange encounters will be favored when the incoming single star is heavier than one of the members of the binary, such that the modulus of the final binding energy is higher than in the former binary.

In this work we investigate how the dynamical properties of GCs affect their $\gamma$-ray and X-ray emission. Our main goal is to test whether the stellar encounters (represented quantitatively by $\Gamma$) represent the only formation channel for MSPs in GCs and what is the impact of secondary encounters once a compact binary is formed, i.e., do they act in favor of ionization or hardening of the formed compact binaries? Furthermore, we look for extended $\gamma$-ray emission in the GCs and investigate if their spectra is indeed consistent with populations of MSPs. 

This paper is organized as follows. We describe the criteria for selecting our GC sample and collected archival data in \S \ref{sec:sample}, and detail the new observations and analyses in \S \ref{sec:observations}. The results are presented in \S \ref{sec:resultados}, while we discuss and summarize our findings in \S \ref{sec:discussion} and \S \ref{sec:conclusions}.

\section{Sample and data description}
\label{sec:sample}

Our sample consists of 124 Galactic GCs listed in \cite{bahramian2013stellar} for which the stellar encounter rates have been precisely measured. The computation of $\Gamma$ by these authors is based on Eq. \ref{eq:enc_rate} and does not rely on the assumption that the clusters' luminosity profile follows the King model \citep{king1962structure,king1966structureIII}, i.e. core-collapsed and non-core-collapsed GCs can be treated in the same way. Furthermore, based on Monte Carlo simulations, they provide asymmetrical error estimates in $\Gamma$ for each GC in the sample.

Other parameters, such as the distance from the Sun, core radii, central velocity dispersion, core density, right ascension, and declination have been taken from the 3rd globular cluster database\footnote{\url{https://people.smp.uq.edu.au/HolgerBaumgardt/globular/}}, which is a compilation of observational and theoretical results for Galactic GCs based mainly on the works by \cite{Baumgardt2017n-body,Sollima2017global,Baumgardt2018catalogue} and \cite{Vasiliev2021Gaia}. Information about radio pulsars and MSPs in GCs have been taken from the online database for pulsars in GCs\footnote{\url{http://www.naic.edu/~pfreire/GCpsr.html}}.

We found archival high-resolution Chandra X-ray observations for 72 out of the 124 clusters in our sample, however, some of these observations are not deep enough for our purposes (see \S \ref{sec:observations} for details). For each GC, we download data from the Chandra Source Catalog \citep[CSC2;][]{evans2010_CSC,evans2020_CSC2} within a circular region contained by the cluster's half-light radius (ranging from $0.3'$ to $4.8'$ in our sample), then having access to the total number of X-ray sources, the X-ray flux of each source, and the flux sensitivity limit of the given observation.

Finally, we use nearly 14 years of \textit{Fermi}-LAT \citep{atwood2009_Fermi-LAT} observations collected from 4th August 2008 up to 10th June 2022 (mission elapsed time from 239557417 to 676512005) to look for $\gamma$-ray emission from the 124 GCs in our sample. We use PASS8 \citep{atwood2013pass} R3 \citep{bruel2018_PASS8R3} data in the energy range 0.3 -- 300 GeV for regions of interest (RoIs) of $14^{\circ} \times 14^{\circ}$, $0.1^{\circ}$ per pixel, centered in each GC. Since our sample of 124 GCs does not cover all GCs in the Galaxy, there are 4 GCs listed in 4FGL-DR3 that do not appear in our sample, namely: 2MASS-GC01, GLIMPSE C01, GLIMPSE C02, and NGC 6838, which are kept out of our analysis. We also exclude from the analysis the GCs NGC 6624, NGC 6626, and NGC 6652, since their emissions are dominated by single pulsars \citep{freire2011fermi,johnson2013broadband,gautam2022pulsarinNGC6652}. Furthermore, we treat the $\gamma$-ray observations of Terzan 1, listed in 4FGL-DR3 as a GC, as upper limits since its optical center lies outside the 99\% confidence region of the closest $\gamma$-ray source \citep[some inconsistencies with the association of Terzan 1 with a $\gamma$-ray source have also been found by][]{wu2022gamma}. Detailed information on the \textit{Fermi}-LAT analysis is given in \S \ref{sec:observations}.

\section{Data analyses}
\label{sec:observations}

\subsection{\textit{Chandra}}

To investigate how the stellar encounter rate affects the formation of compact binary systems in GCs, we simply compare $\Gamma$ with the total number of X-ray sources, $N_{\mathrm{x}}$, found within the half-light radii of GCs (see \S \ref{sec:sample} for data acquisition details) after subtracting the estimated number of background sources based on the following $\log N-\log S$ relation \citep{giacconi2001first}:

\begin{equation}
    N(>S) = 1200\left(\frac{S}{2\times 10^{-15}}   \right)^{-1.0 \pm 0.2} \rm{sources~deg^{-2}},
\end{equation}
where $N$ is the expected number of background sources per square degree and $S$ is the flux sensitivity of the observation in a given region. 

Since the observations performed with \textit{Chandra} do not necessarily have the same exposure times, we have a non-homogeneous X-ray sample of the central region of GCs. To fix this, we set a luminosity limit of $L_{\mathrm{x,lim}} = 2\times 10^{31}$ erg s$^{-1}$, i.e. we only use GCs for which the X-ray observations can reveal sources with luminosities greater than $2\times 10^{31}$ erg s$^{-1}$. This reduces our X-ray sample from 72 to 38 GCs. To check if the choice of $L_{\mathrm{x,lim}}$ could significantly impact the results, we repeated exactly the same analysis using other values for $L_{\mathrm{x,lim}}$ and in the end they all give similar results for the relation $N_{\mathrm{x}} \times \Gamma$, although always with higher variances. In summary, if we adopt a limit that is too low (e.g. $L_{\mathrm{x,lim}} =  4\times 10^{30}$ erg s$^{-1}$), only a few GCs with very deep observations can be used; on the other hand, if the luminosity limit is too high (e.g. $L_{\mathrm{x,lim}} = 10^{32}$ erg s$^{-1}$) we can use more GCs, but with fewer sources detected in each one of them. The final choice of $L_{\mathrm{x,lim}} = 2\times 10^{31}$ erg s$^{-1}$ is the one that minimizes the variance of the $N_{\mathrm{x}} \times \Gamma$ relation described in \S \ref{sec:resultados}. In this analysis, we also discard GCs for which the final number of X-ray sources is less than 2, further reducing our X-ray sample to 30 GCs.

\subsection{\textit{Fermi}-LAT}
\label{sec:analysis-fermi}

The \textit{Fermi}-LAT data analysis is performed with \texttt{Fermitools}\footnote{\url{https://fermi.gsfc.nasa.gov/ssc/data/analysis/software/}} v2.2.0, \texttt{Fermipy}\footnote{\url{https://fermipy.readthedocs.io/en/latest/index.html}} v1.2 \citep{wood2017fermipy}, and \texttt{easyFermi}\footnote{\url{https://github.com/ranieremenezes/easyFermi}} v1.0.10 \citep{deMenezes2022easyfermi} by means of a binned likelihood analysis and using MINUIT as minimizer. For modeling each RoI, we consider the $\gamma$-ray sources listed in 4FGL-DR3 as well as all sources found with the \texttt{Fermipy} function \texttt{find\_sources()} presenting a statistical significance above $8\sigma$. In the fit, the normalization of all sources lying within a radius of $5^{\circ}$ from the center of the RoIs is left free to vary, while the spectral shape parameters are left free only if the source presents a significance higher than $15\sigma$ over the $\sim 14$ years of \textit{Fermi}-LAT observations. Furthermore, we also include in the model those sources listed in 4FGL-DR3 and lying up to $4^{\circ}$ outside the RoIs.

We divide the data into 8 logarithmically spaced bins per energy decade (in the range 0.3 -- 300 GeV) and use \texttt{SOURCE} events ($evclass=128$) detected in the front or back layers of the tracker ($evtype=3$). The data are also filtered for good time intervals with \texttt{DATA\_QUAL} $> 0$ and the recommended instrument configuration for science \texttt{LAT\_CONFIG} $== 1$, while the maximum zenith angle cut is $\theta_z = 95^{\circ}$ for photons with energies $\leq 1$ GeV and $\theta_z = 105^{\circ}$ above 1 GeV. Background emission from the Milky Way and extragalactic isotropic component are modeled with the interstellar emission model\footnote{\url{https://fermi.gsfc.nasa.gov/ssc/data/access/lat/BackgroundModels.html}} \textit{gll\_iem\_v07} and the isotropic spectral template \textit{iso\_P8R3\_SOURCE\_V3\_v1}.

The significance of our observations is set by a test statistic (TS) defined as $TS = 2(\mathcal{L}_1-\mathcal{L}_0)$, where the term inside parentheses is the difference between the maximum log-likelihoods with ($\mathcal{L}_1$) and without ($\mathcal{L}_0$) including our target in the model. The corresponding significance, in $\sigma$, is approximated by $\sqrt{TS}$ \citep{mattox1996likelihood} in the case of one degree of freedom. Throughout this work, we adopt two spectral models for our targets, both of which have been used in the past to model pulsars and GCs \citep{abdollahi2020_4FGL,abdo2013_2PC}: a power-law with exponential cutoff (PLEC),

\begin{equation}
    \frac{dN}{dE} = N_0\left(\frac{E}{E_0}\right)^{-\alpha}e^{-E/E_{\mathrm{c}}}, 
    \label{eq:PLEC}
\end{equation}
where $N_0$ is the normalization, $\alpha$ is the photon index, $E_0 \equiv 1000$ is the energy scale in MeV, and $E_{\mathrm{c}}$ is the energy cutoff; and a log-parabola

\begin{equation}
    \frac{dN}{dE} = N_0\left(\frac{E}{E_0}\right)^{-(\alpha + \beta\log(E/E_0))},
    \label{eq:LogPar}
\end{equation}
where $\beta$ is the spectral curvature parameter.


\subsubsection{Extended emission}
\label{subsec:extended}

The dynamical friction acting on the stars of a GC makes the heavier objects, such as neutron stars and binary systems, sink to the GC's core \citep{chandrasekhar1943dynamical}. We then expect that close or exchange encounters involving these neutron stars will form compact binary systems that can eventually fill the core of GCs with MSPs. 

Since the largest core radius in our sample is $r_{\mathrm{c}} = 2.7$ arcmin (belonging to NGC 5139/$\Omega$ Centauri), we expect the cumulative emission from MSPs to appear as point-like to \textit{Fermi}-LAT up to a few GeV. On the other hand, relativistic leptons escaping the magnetospheres of MSPs can propagate within GCs and scatter soft stellar photons to the $\gamma$-ray energy range \citep{bednarek2007high,zajczyk2013numerical,bednarek2016tev}. If these leptons are able to leave the cores of GCs before interacting with the soft photons, it would be possible to detect extended $\gamma$-ray emission from GCs. Measuring this extended emission may help us to investigate at least two different phenomena: i) how the neutron stars sink to the core of GCs with distinct dynamical properties, and ii) how relativistic leptons propagate in the interior of GCs.

We look for extended $\gamma$-ray emission from the 5 GCs with the largest optical half-light radii (HLRs) detected with the \textit{Fermi}-LAT, namely NGC 104, NGC 5139, NGC 6656, NGC 6397, and NGC 6752. To improve the resolution of the LAT data with respect to the analysis described in \S \ref{sec:analysis-fermi}, this time we only use photons with more than 1 GeV, and a bin size of $0.02^{\circ}$ (instead of $0.1^{\circ}$). To assess the significance of these results, we build a control sample by repeating the analysis for 100 blazars, which are expected to be point-like. These blazars are listed in 4FGL-DR3 and have fluxes ($> 1$ GeV) and Galactic latitudes similar to those of the 5 GCs tested here (more details in \S \ref{sec:results_extended}). The best-fit extension is found by performing a likelihood profile scan over the source width (68\% containment) in 15 equally spaced width intervals from $0.02^{\circ}$ up to $0.3^{\circ}$, and fitting for the extension that maximizes the likelihood of the adopted models, which are a 2D radial Gaussian and a disk. The results are shown in \S \ref{sec:results_extended}.

\section{Results}
\label{sec:resultados}

The results from the RoI fits performed in the \textit{Fermi}-LAT data are shown in Table \ref{tab:fits_results}. We find a total of 39 GCs with TS $> 16$, seven of which are not listed as GCs in 4FGL-DR3: NGC 6380, NGC 6517, and NGC 6723, which have no counterpart in 4FGL-DR3; Terzan 6, which is associated in 4FGL-DR3 with the unknown X-ray source 1RXS J175042.7-310312; and NGC 6342, NGC 6528, and NGC 6637, lying within the error ellipse of the unidentified $\gamma$-ray sources (UGSs) 4FGL J1720.8-1937, 4FGL J1804.9-3001, and 4FGL J1830.7-3219, respectively. Except for the 5 GCs mentioned in the last paragraph of \S \ref{sec:sample} (i.e. Terzan 1 and the 4 GCs not included in our analysis due to the absence of a reliable value for the stellar encounter rate), all GCs listed in 4FGL-DR3 are also detected in this work.

\begin{table*}
    \centering
    \begin{tabular}{l|c|c|c|c|c|c|c|c}
     Name & $\Gamma$ & $N_{\mathrm{P}}$ & TS & eFlux & Norm & $\alpha$ & $E_{\mathrm{c}}$ & $\beta$ \\
  &   &   &   & $10^{-12}$ erg cm$^{-2}$ s$^{-1}$ & $10^{-13}$ cm$^{-2}$ s$^{-1}$ MeV$^{-1}$ &   & GeV &   \\
\hline
NGC  104  & $1000.0^{+ 154.0 }_{- 134.0 }$ &  29  &  12124.76  & $ 23.86 \pm  0.45 $ & $  73.10 \pm  2.70 $ & $ 1.32 \pm  0.05 $ & $ 2.77 \pm  0.20 $ & -- \\
NGC  1851  & $1530.0^{+ 198.0 }_{- 186.0 }$ &  15  &  65.66  & $ 1.12 \pm  0.20 $ & $ 2.39 \pm  0.43 $ & $ 2.30 \pm  0.24 $ & --  & $ 0.14 \pm  0.08 $ \\
NGC  1904  & $116.0^{+ 67.6 }_{- 44.7 }$ &  0  &  66.41  & $ 1.41 \pm  0.22 $ & $  3.55 \pm  0.82 $ & $ 2.28 \pm  0.34 $ & $ 7.23 \pm  6.86 $ & -- \\
NGC  2808  & $923.0^{+ 67.2 }_{- 82.7 }$ &  0  &  196.68  & $ 2.95 \pm  0.27 $ & $  8.58 \pm  1.41 $ & $ 1.88 \pm  0.23 $ & $ 4.04 \pm  1.73 $ & -- \\
NGC  362  & $735.0^{+ 137.0 }_{- 117.0 }$ &  5  &  42.16  & $ 0.80 \pm  0.18 $ & $  1.88 \pm  0.56 $ & $ 1.73 \pm  0.56 $ & $ 5.48 \pm  5.34 $ & -- \\
NGC  5139  & $90.4^{+ 26.3 }_{- 20.4 }$ &  18  &  1423.60  & $ 10.24 \pm  0.41 $ & $  41.05 \pm  4.00 $ & $ 1.12 \pm  0.14 $ & $ 1.92 \pm  0.28 $ & -- \\
NGC  5286  & $458.0^{+ 58.4 }_{- 60.7 }$ &  0  &  31.92  & $ 1.57 \pm  0.33 $ & $  3.52 \pm  0.90 $ & $ 2.28 \pm  0.37 $ & $ 11.67 \pm  8.58 $ & -- \\
NGC  5904  & $164.0^{+ 38.6 }_{- 30.4 }$ &  7  &  61.57  & $ 1.12 \pm  0.29 $ & $ 2.77 \pm  0.88 $ & $ 1.35 \pm  0.88 $ & --  & $ 0.65 \pm  0.49 $ \\
NGC  6093  & $532.0^{+ 59.1 }_{- 68.8 }$ &  0  &  127.40  & $ 3.15 \pm  0.38 $ & $  6.36 \pm  1.26 $ & $ 1.54 \pm  0.19 $ & $ 5.83 \pm  1.76 $ & -- \\
NGC  6139  & $307.0^{+ 95.4 }_{- 82.1 }$ &  0  &  101.24  & $ 3.66 \pm  0.51 $ & $  11.51 \pm  2.59 $ & $ 1.41 \pm  0.42 $ & $ 2.82 \pm  1.32 $ & -- \\
NGC 6205  & $68.9^{+ 18.1 }_{- 14.6 }$ &  6  &  19.06  & $ 0.37 \pm  0.12 $ & $  0.46 \pm  0.34 $ & $1.00$* & -- & $ 0.73 \pm  0.44 $ \\
NGC  6218  & $13.0^{+ 5.44 }_{- 4.03 }$ &  2  &  49.33  & $ 1.14 \pm  0.33 $ & $  1.69 \pm  0.30 $ & $ 0.97 \pm  0.12 $ & $ 5.10 \pm  1.01 $ & -- \\
NGC  6266  & $1670.0^{+ 709.0 }_{- 569.0 }$ &  9  &  1727.84  & $ 16.71 \pm  0.60 $ & $  52.16 \pm  3.47 $ & $ 1.41 \pm  0.11 $ & $ 2.84 \pm  0.38 $ & -- \\
NGC  6304  & $123.0^{+ 53.8 }_{- 22.0 }$ &  0  &  49.30  & $ 2.12 \pm  0.39 $ & $  8.95 \pm  0.63 $ & $ 0.37 \pm  0.28 $ & $ 1.41 \pm  0.11 $ & -- \\
NGC  6316  & $77.0^{+ 25.4 }_{- 14.8 }$ &  0  &  361.57  & $ 8.76 \pm  0.62 $ & $  25.25 \pm  2.87 $ & $ 1.54 \pm  0.19 $ & $ 3.41 \pm  0.87 $ & -- \\
NGC  6341**  & $270.0^{+ 30.1 }_{- 29.0 }$ &  1  &  111.87  & $ 1.50 \pm  0.18 $ & $  6.47 \pm  2.15 $ & $ 1.86 \pm  0.39 $ & $ 2.00 \pm  1.01 $ & -- \\
NGC  6342  & $44.8^{+ 14.4 }_{- 12.5 }$ &  2  &  55.15  & $ 2.79 \pm  0.49 $ & $  2.96 \pm  2.90 $ & $ 2.15 \pm  0.26 $ & $ 3.00 \pm  0.24 $ & -- \\
NGC  6380  & $116.0^{+ 19.1 }_{- 14.2 }$ &  0  &  31.12  & $ 2.13 \pm  0.42 $ & $ 4.47 \pm  1.29 $ & $ 0.93 \pm  0.27 $ & --  & $ 0.75 \pm  0.01 $ \\
NGC  6388  & $899.0^{+ 238.0 }_{- 213.0 }$ &  0  &  1628.64  & $ 15.47 \pm  0.52 $ & $  50.19 \pm  3.24 $ & $ 1.49 \pm  0.08 $ & $ 2.82 \pm  0.31 $ & -- \\
NGC  6397  & $84.1^{+ 18.3 }_{- 18.3 }$ &  2  &  56.82  & $ 1.84 \pm  0.30 $ & $ 5.00 \pm  0.89 $ & $ 2.28 \pm  0.25 $ & --  & $ 0.40 \pm  0.19 $ \\
NGC  6402  & $124.0^{+ 31.8 }_{- 30.2 }$ &  5  &  85.64  & $ 3.09 \pm  0.39 $ & $  9.06 \pm  1.92 $ & $ 1.77 \pm  0.22 $ & $ 3.78 \pm  1.41 $ & -- \\
NGC  6440  & $1400.0^{+ 628.0 }_{- 477.0 }$ &  8  &  455.73  & $ 12.10 \pm  0.74 $ & $ 27.33 \pm  1.79 $ & $ 2.13 \pm  0.10 $ & --  & $ 0.23 \pm  0.06 $ \\
NGC  6441  & $2300.0^{+ 974.0 }_{- 635.0 }$ &  9  &  760.85  & $ 12.22 \pm  0.53 $ & $ 28.70 \pm  1.31 $ & $ 2.13 \pm  0.05 $ & --  & $ 0.26 \pm  0.01 $ \\
NGC  6517  & $338.0^{+ 152.0 }_{- 97.5 }$ &  17  &  23.34  & $ 2.09 \pm  0.48 $ & $ 4.80 \pm  1.35 $ & $ 1.75 \pm  0.25 $ & --  & $ 0.38 \pm  0.01 $ \\
NGC  6528  & $278.0^{+ 114.0 }_{- 49.5 }$ &  0  &  48.97  & $ 3.60 \pm  0.55 $ & $  2.22 \pm  1.67 $ & $ 2.01 \pm  0.11 $ & $ 3.58 \pm  1.17 $ & -- \\
NGC  6541  & $386.0^{+ 95.2 }_{- 63.1 }$ &  0  &  156.16  & $ 3.14 \pm  0.31 $ & $  9.15 \pm  1.79 $ & $ 1.42 \pm  0.21 $ & $ 3.12 \pm  0.94 $ & -- \\
NGC  6637  & $89.9^{+ 36.0 }_{- 18.1 }$ &  0  &  38.60  & $ 1.82 \pm  0.41 $ & $ 3.70 \pm  0.92 $ & $ 1.96 \pm  0.39 $ & --  & $ 0.21 \pm  0.19 $ \\
NGC  6652  & $700.0^{+ 292.0 }_{- 189.0 }$ &  2  &  180.10  & $ 3.77 \pm  0.34 $ & $  11.78 \pm  2.03 $ & $ 1.71 \pm  0.20 $ & $ 3.30 \pm  1.07 $ & -- \\
NGC  6656  & $77.5^{+ 31.9 }_{- 25.9 }$ &  4  &  143.75  & $ 3.92 \pm  0.31 $ & $  25.83 \pm  1.04 $ & $ 1.09 \pm  0.55 $ & $ 1.22 \pm  0.08 $ & -- \\
NGC  6681  & $1040.0^{+ 267.0 }_{- 192.0 }$ &  0  &  27.29  & $ 1.10 \pm  0.26 $ & $ 2.05 \pm  0.89 $ & $ 1.23 \pm  0.39 $ & --  & $ 0.49 \pm  0.15 $ \\
NGC  6712  & $30.8^{+ 5.63 }_{- 6.64 }$ &  1  &  35.37  & $ 2.59 \pm  0.62 $ & $ 6.71 \pm  1.55 $ & $ 2.34 \pm  0.38 $ & --  & $ 0.32 \pm  0.31 $ \\
NGC  6715  & $2520.0^{+ 226.0 }_{- 274.0 }$ &  0  &  20.46  & $ 1.17 \pm  0.29 $ & $  3.28 \pm  1.26 $ & $ 1.69 \pm  0.40 $ & $ 3.89 \pm  2.33 $ & -- \\
NGC  6717  & $39.8^{+ 21.8 }_{- 13.7 }$ &  0  &  121.49  & $ 3.00 \pm  0.43 $ & $  10.46 \pm  2.66 $ & $ 1.25 \pm  0.44 $ & $ 2.25 \pm  1.00 $ & -- \\
NGC  6723  & $11.4^{+ 8.01 }_{- 4.39 }$ &  0  &  25.42  & $ 1.25 \pm  0.34 $ & $ 2.72 \pm  0.78 $ & $ 2.01 \pm  0.44 $ & --  & $ 0.26 \pm  0.23 $ \\
NGC  6752  & $401.0^{+ 182.0 }_{- 126.0 }$ &  9  &  384.60  & $ 3.34 \pm  0.25 $ & $  15.71 \pm  2.77 $ & $ 1.17 \pm  0.27 $ & $ 1.67 \pm  0.40 $ & -- \\
NGC  7078  & $4510.0^{+ 1360.0 }_{- 986.0 }$ &  9  &  118.00  & $ 2.01 \pm  0.27 $ & $  20.67 \pm  0.41 $ & $ 0.92 \pm  0.55 $ & $ 0.85 \pm  0.05 $ & -- \\
Terzan  2  & $22.1^{+ 28.6 }_{- 14.4 }$ &  0  &  96.22  & $ 5.22 \pm  1.09 $ & $ 15.36 \pm  0.53 $ & $ 0.93 \pm  0.56 $ & $ 2.34 \pm  0.41 $ & -- \\
Terzan  5  & $6800.0^{+ 1040.0 }_{- 3020.0 }$ &  42  &  6631.77  & $ 70.53 \pm  1.20 $ & $ 194.64 \pm  5.64 $ & $ 1.67 \pm  0.04 $ & $ 3.94 \pm  0.26 $ & -- \\
Terzan  6  & $2470.0^{+ 5070.0 }_{- 1720.0 }$ &  0  &  80.18  & $ 6.38 \pm  0.72 $ & $ 13.12 \pm  3.01 $ & $ 0.90 \pm  0.48 $ & -- & $ 1.10 \pm  0.30 $ 
    \end{tabular}
    \caption{The 39 GCs in our sample detected by \textit{Fermi}-LAT with TS $> 16$. The columns represent the GC name, two-body encounter rate $\Gamma$ (normalized such that $\Gamma_{\mathrm{NGC104}} = 1000$), number of pulsars detected in radio $N_{\mathrm{P}}$ (as of January 2023), TS, integrated energy flux from 0.3 to 300 GeV, normalization and curvature parameters of the spectral models described by Eqs. \ref{eq:PLEC} and \ref{eq:LogPar}. The ``*'' indicates a fixed parameter. The ``**'' indicates a GC for which the $\gamma$-ray emission is possibly dominated by a single MSP \citep{zhang2023arXiv_likely}.}
    \label{tab:fits_results}
\end{table*}

Most of the GCs in our sample are well described by a PLEC spectrum (Eq. \ref{eq:PLEC}), however, for 13 GCs, we adopt a log-parabola (Eq. \ref{eq:LogPar}) model. We favor the log-parabola model only when the fitting of the PLEC spectrum is not possible (typically for targets with TS $< 80$) or when the significance of the log-parabola model is at least $2\sigma$ higher than that achieved for the PLEC model (this is true only for NGC 6440 and NGC 6441).

\begin{figure*}
    \centering
    \includegraphics[width=\linewidth]{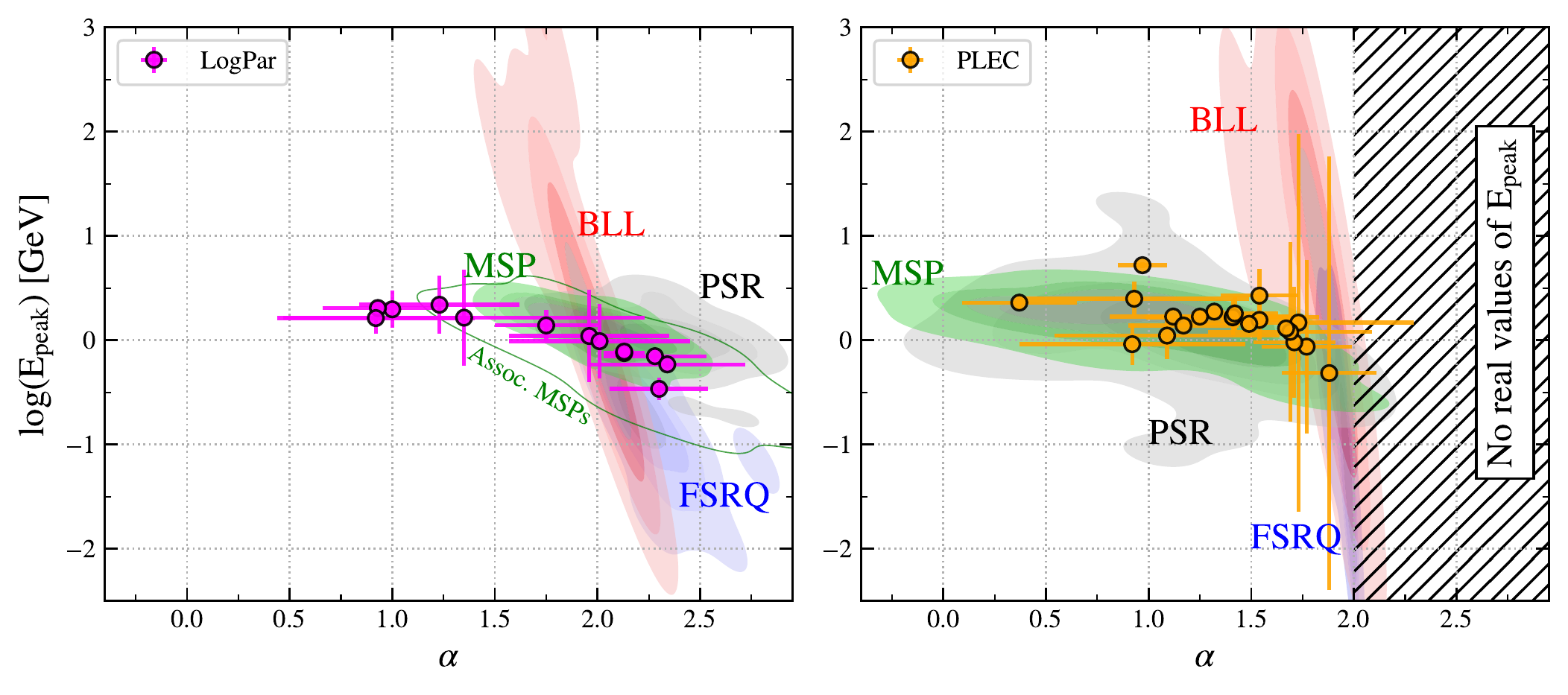}
    \caption{The distribution of $\gamma$-ray GCs in the $E_{\mathrm{peak}} \times \alpha$ space for the log-parabola (left) and PLEC (right) models. Each colored region represents the distributions of $E_{\mathrm{peak}}$ and $\alpha$ values of a specific astrophysical class listed in 4FGL, where PSR stands for identified young pulsars, MSP for identified millisecond pulsars, FSRQ for identified flat spectrum radio quasars, and BLL for identified BL Lacs. The green solid line on the left panel represents the region occupied by associated MSPs, which is substantially broader than the region occupied by the identified MSPs, and is consistent with all data points within their error bars. The orange and magenta points represent the GCs detected with \textit{Fermi}-LAT in this work for which $E_{\mathrm{peak}} > 0.3$ GeV (i.e. the minimum energy set in \S \ref{sec:sample}). As expected, the spectral features of GCs resemble those of MSPs.}
    \label{fig:colorcolor}
\end{figure*}

The results shown in Table \ref{tab:fits_results} allow us to quickly investigate the spectra of the GCs and to check if they are consistent with a population of MSPs. In Fig. \ref{fig:colorcolor} we summarize the results of Table \ref{tab:fits_results} by comparing the distributions of spectral energy peak ($E_{\mathrm{peak}}$) and spectral index ($\alpha$) of the GCs with those of different populations listed in 4FGL. The values for $E_{\mathrm{peak}}$ are found by multiplying Eqs. \ref{eq:PLEC} and \ref{eq:LogPar} by E$^2$ and setting their first derivatives to zero, such that $E_{\mathrm{peak}} = E_{\mathrm{c}}(2+\alpha)$ for the PLEC model, and $E_{\mathrm{peak}} = E_0e^{(2-\alpha)/2\beta}$ for the log-parabola one. On the left panel of Fig. \ref{fig:colorcolor}, we identify the regions corresponding to different classes of astrophysical sources in the log-parabola $E_{\mathrm{peak}} \times \alpha$ space by simply collecting these parameters from high-significance ($>30\sigma$) identified $\gamma$-ray sources listed in 4FGL-DR3 (i.e. those sources for which the association with a low-energy counterpart is guaranteed\footnote{As defined in the 4FGL paper \citep{abdollahi2020_4FGL}, a $\gamma$-ray source is considered as \textit{identified} based on pulsations, correlated variability or correlated angular sizes with observations at other wavelengths. On the other hand, a $\gamma$-ray source is considered \textit{associated} when an association probability is attributed to it based mainly on spatial coincidence and relative radio or X-ray brightness.}). These regions are shown in different colors together with the data points for the GCs and represent the typical spectral parameters of these populations averaged over 12 years of \textit{Fermi}-LAT observations. We repeat this process on the right panel of Fig. \ref{fig:colorcolor} for the PLEC model, although this time we adopt 4FGL-DR2 \citep{ballet2020_4FGL-DR2} because it uses a PLEC model similar to the one defined in Eq. \ref{eq:PLEC}, except for the exponential factor, which is given by $aE^b$ instead of $E/E_{\mathrm{c}}$, where $a$ and $b$ are constants. Since 4FGL-DR2 doesn't provide a value for $E_{\mathrm{peak}}$, we compute it ourselves as $E_{\mathrm{peak}} = \sqrt[b]{(2-\alpha)/ab}$ (note that there are no real values for $E_{\mathrm{peak}}$ if $\alpha > 2$). 

Almost all GCs lie in regions consistent with MSPs, leaving little room for false-positive detections, as e.g. due to a background blazar. Among the five outliers in the left panel of Figure \ref{fig:colorcolor}, one of them (i.e. NGC 6205) was forced to have $\alpha = 1$ in our fit due to low statistics, as shown in Table \ref{tab:fits_results}, while the other four GCs are still consistent with the MSP region in the condition that we build this region using the identified and the associated MSPs listed in 4FGL-DR3 (i.e. the green region in the left panel of Figure \ref{fig:colorcolor} gets broader). Other exotic possibilities for the interpretation of the $\gamma$-ray emission from GCs can be found in literature \citep{feng2012search,brown2018understanding,evans2022gamma,evans2022dark}, but they are beyond the scope of this work. For the GCs with TS $<16$, we provide the integrated energy flux (from 0.3 to 300 GeV) upper limits in Appendix \ref{appendix}.

\subsection{The formation of compact binary systems}

\begin{figure}
    \centering
    \includegraphics[width=\linewidth]{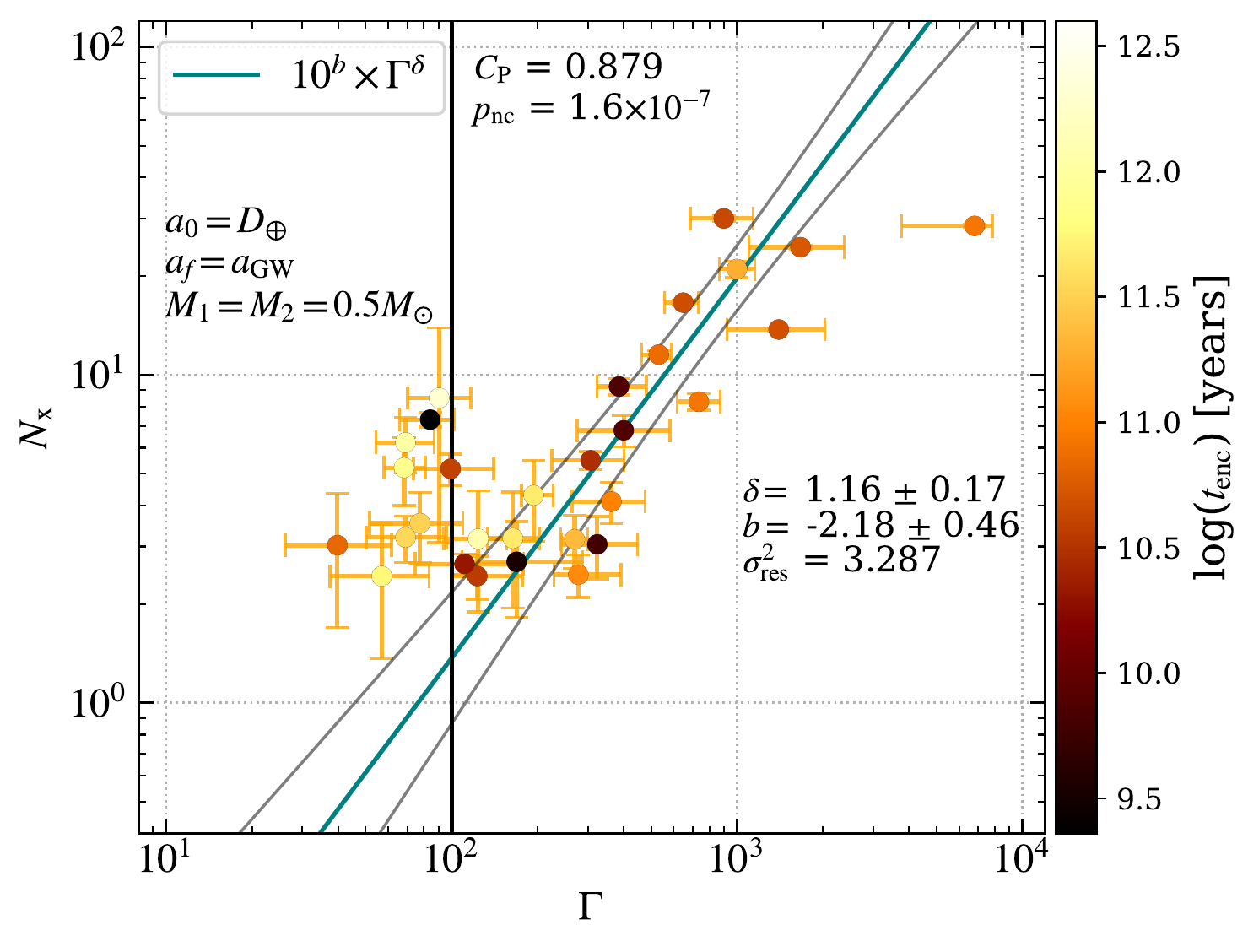}
    \caption{Correlation between the total number of X-ray sources and the stellar encounter rate $\Gamma$. The vertical black line denotes 10\% of the value of $\Gamma_{\mathrm{NGC104}}$. The color bar represents the timescale ($t_{\mathrm{enc}}$) that a primordial binary, where the stars have masses $M_1 = M_2 = 0.5$ M$_{\odot}$, takes to shrink from $D_{\oplus}$ until the separation where the loss of energy via gravitational waves becomes relevant, $a_{\mathrm{GW}}$.}
    \label{fig:X-ray-EncRate}
\end{figure}

If compact binaries dominate the X-ray emission in GCs and are mainly formed in stellar close encounters, we must see a linear correlation between the total number of X-ray sources, $N_x$, and the stellar encounter rate $\Gamma$. In Fig. \ref{fig:X-ray-EncRate} we see that, for high values of $\Gamma$ (i.e. $\Gamma \gtrsim$ 100, or 10\% the value of $\Gamma$ for NGC 104), this correlation is indeed linear within the errors of the fit. The fit here is performed with an orthogonal distance regression since we have error bars in both variables $N_x$ and $\Gamma$, and includes only GCs with $\Gamma > 100$. The Pearson coefficient for linear correlation, $C_{\mathrm{P}} = 0.879$, is quite high, indicating that these variables are indeed correlated. Furthermore, the p-value for non-correlation, $p_{\mathrm{nc}} = 1.6 \times 10^{-7}$, is very small. If instead we use all GCs when performing the fit, the Pearson correlation coefficient decreases to $C_{\mathrm{P}} = 0.759$ (which is still a relatively high value) and the p-value for non-correlation becomes $p_{\mathrm{nc}} = 1.15 \times 10^{-6}$.

For GCs with $\Gamma < 100$, we see an increase in the total number of X-ray sources with respect to what is expected from the fit. This may be a hint that, for GCs where stellar encounters are not that frequent, there must be another mechanism forming compact binaries, or even that low values of $\Gamma$ in a GC favor another kind of X-ray emitting astrophysical process unrelated to the compact binaries. One of the possibilities that we investigate is if primordial binaries, made of two stars with equal masses $M_1 = M_2 = 0.5$ M$_{\odot}$ and separated by a distance of 1 astronomical unity (herein $D_{\oplus}$), born in the GCs have enough time to shrink until becoming a compact binary (i.e. until reaching $a_{\mathrm{GW}}$, the orbital separation at which the energy loss via gravitational waves is higher than the loss via stellar encounters. More details in \S \ref{sec:subsequent_encounters}). This shrinking time, $t_{\mathrm{enc}}$, is represented by the color bar in Fig. \ref{fig:X-ray-EncRate} and its detailed calculation is shown in \S \ref{sec:subsequent_encounters}. The exact values for the shrinking times here are not very important, since we choose quite generic parameters for the test binary system, however, they give us an idea of which GCs require the longest time for a primordial binary to shrink. We see that the primordial binaries in GCs with $\Gamma < 100$ take, on average, longer times to shrink than the binaries in GCs with $\Gamma > 100$. We therefore conclude that, whatever the mechanism generating these X-ray sources, it is unlikely that they are compact binaries formed by either close stellar encounters or the shrinkage of primordial binaries. 

\begin{figure*}
    \centering
    \includegraphics[width=\linewidth]{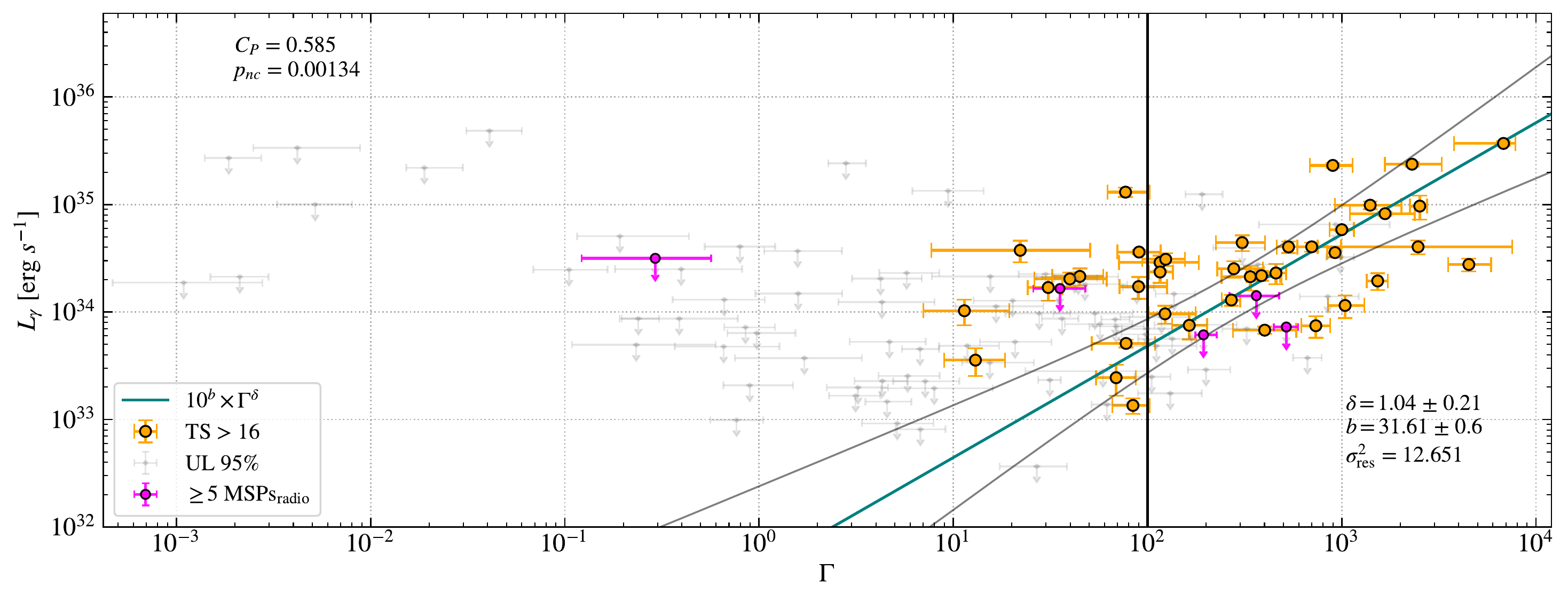}
    \caption{Correlation between the $\gamma$-ray luminosity $L_{\gamma}$ and the stellar encounter rate $\Gamma$. The orange points represent the GCs with TS $> 16$, the magenta points represent the 95\% $\gamma$-ray upper limits for GCs with 5 or more MSPs detected in radio, and the gray points represent the 95\% $\gamma$-ray upper limits for all other GCs in our sample. The vertical black line denotes 10\% of the value of $\Gamma_{\mathrm{NGC104}}$.}
    \label{fig:LgxGamma}
\end{figure*}

We repeat this analysis with $\gamma$-ray data, but this time using the isotropic luminosity, $L_{\gamma}$, of the GCs instead of counting sources, since the resolution of \textit{Fermi}-LAT does not allow for it. In Fig. \ref{fig:LgxGamma} we see that there is also a linear correlation between $L_{\gamma}$ and $\Gamma$, although it presents a higher scatter. The Pearson coefficient for linear correlation in this case is $C_{\mathrm{P}} = 0.585$ and the  p-value for non-correlation is $p_{\mathrm{nc}} = 1.34 \times 10^{-3}$, indicating a weaker correlation with respect to what we see in Fig. \ref{fig:X-ray-EncRate}. The higher scatter in this fit may be explained by whether or not each cluster has been able to form a high-$\dot{E}$ MSP (i.e. $\dot{E} = 10^{35} \sim 10^{36}$ erg s$^{-1}$), as observed in a few clusters \citep{freire2011fermi,johnson2013broadband}, which dominate the $\gamma$-ray emission. In Fig. \ref{fig:LgxGamma} we also see that three of the upper limits breaking the fit belong to GCs with at least 5 MSPs detected in radio (magenta points), indicating that there are $\gamma$-ray sources in these GCs but their $\gamma$-ray emission is too weak to be detected by LAT. The fit here is again performed with an orthogonal distance regression and, based on the X-ray analysis described in the previous paragraphs, we include in the fit only the GCs with $\Gamma > 100$. We see some hint of excess $\gamma$-ray emission for GCs with $\Gamma < 100$, although it is not as evident as the excess observed in Fig. \ref{fig:X-ray-EncRate}. If instead we use all GCs in the fit, the Pearson coefficient slightly decreases to $C_{\mathrm{P}} = 0.55$ and the p-value for non-correlation becomes $p_{\mathrm{nc}} = 2.9 \times 10^{-4}$, indicating that the $L_{\gamma} \times \Gamma$ correlation found here is similar but a bit weaker than those found by \cite{hui2010fundamental} and \cite{deMenezes2019milky}, although these authors used a different approach for computing $\Gamma$.

The high scatter observed for the $L_{\gamma} \times \Gamma$ correlation may also mean that in some GCs the compact binaries do not evolve or had no time to evolve into MSPs. For instance, if the fraction of white dwarfs with respect to the total number of neutron stars in a GC is too high (as a consequence of the stellar initial mass function of the GC), the compact binaries may not evolve into $\gamma$-ray MSPs but could still be bright X-ray sources (in some cases, however, the white dwarf will reach the conditions for collapsing into a neutron star via electron-capture supernovae, as discussed in \S \ref{sec:intro}).

\subsection{Subsequent encounters and binary hardening}
\label{sec:subsequent_encounters}

In the context of the Heggie-Hills law (see \S \ref{sec:intro}), the compact binaries in the core of a GC will get harder after a close encounter if $$\frac{G m_{\mathrm{c}} m_{\star}}{a} > m_{\star} \sigma^2,$$
where $m_{\mathrm{c}}$ is the mass of a compact object, $a$ is the semi-major axis of the binary orbit, $\sigma$ is the velocity dispersion in the core of the GC, and $m_{\star}$ is the mass of a typical star in a GC ($m_{\star} \approx 0.5$ M$_{\odot}$). This means that the binaries will shrink only if 

\begin{equation}
    a < \frac{Gm_{\mathrm{c}}}{\sigma^2}.
    \label{eq:semimajor}
\end{equation}
We can actually rewrite this expression in the context of Eq. \ref{eq:lambda}, since $\Lambda$ depends on $a$. So there is a limit $\Lambda_{lim} = KnGm_{\mathrm{c}}/\sigma^3$ (where $K$ is a constant) at which the binary-single star encounters will tend to ionize the binaries. We then define the term

\begin{equation}
    \lambda = \frac{\Lambda}{\Lambda_{lim}} = \frac{Kna}{\sigma} \frac{\sigma^3}{KnGm_{\mathrm{c}}} = \frac{a\sigma^2}{Gm_{\mathrm{c}}}
    \label{eq:norm_lamb}
\end{equation}
as the normalized encounter rate per formed binary, such that if $\lambda > 1$, the binaries in the GC are typically ionized by stellar encounters. On the other hand, if $\lambda < 1$, the stellar encounters tend to help the binaries to get more and more compact.

Since we know that binaries formed in stellar close encounters must have very compact orbits, with a maximum distance of a few ($a \leq 10$ $R_{\star}$) stellar radii after circularization (see \S \ref{sec:intro}), if we substitute $a \approx 10$ $R_{\star}$ (where $R_{\star} \approx 0.4$ R$_{\odot}$ for a star with $m_{\star} \approx 0.5$ M$_{\odot}$) and assume $m_{\mathrm{c}} = 1.5$ M$_{\odot}$ (as for a typical neutron star) in Eq. \ref{eq:norm_lamb}, we can tell if these encounters are ionizing or shrinking the binaries in the cores of each GC in our sample by simply knowing their central dispersion velocities $\sigma$.

\begin{figure*}
    \centering
    \includegraphics[width=\linewidth]{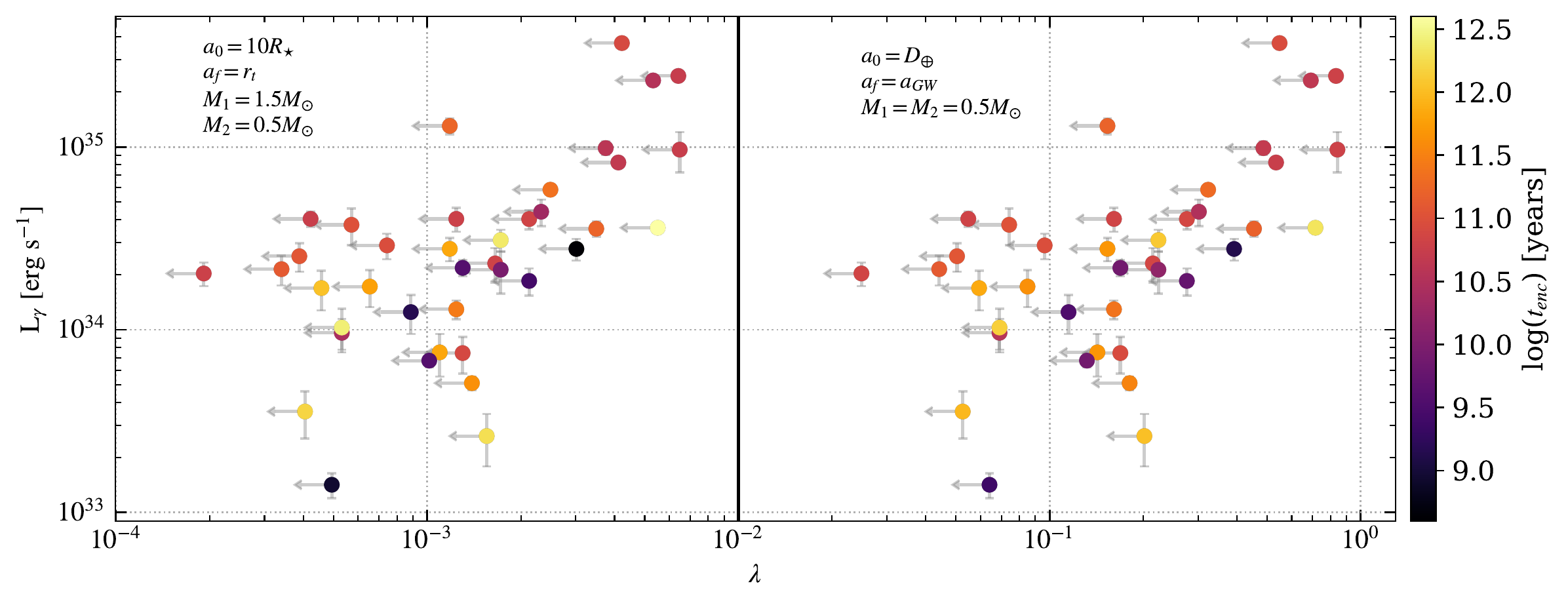}
    \caption{The $\gamma$-ray luminosity vs. the normalized encounter rate per formed binary, $\lambda$. We see that all GCs present $\lambda \ll 1$ for compact binaries (left side), meaning that the secondary stellar encounters will on average make the binaries harder. On the right side of the figure we test the effect of secondary encounters in a primordial binary made of equal-mass stars with 0.5 M$_{\odot}$. We see that even in this case, the secondary encounters tend to make the binaries harder. The color bar gives the shrinking time of these binaries due to secondary encounters: even if these encounters help in shrinking the binary, they are quite inefficient.}
    \label{fig:lambda}
\end{figure*}

In Fig. \ref{fig:lambda}, we show the nuclear values of $\lambda$ in two distinct situations. On the left side of the plot, we compute $\lambda$ for the cores of all GCs detected in $\gamma$ rays assuming that the compact binary systems have been formed in stellar close encounters, where the mass of the compact object is $m_{\mathrm{c}} = 1.5$ M$_{\odot}$, the mass of the main sequence companion star is $m_{\star} = 0.5$ M$_{\odot}$, the central velocity dispersion ($\sigma$) is collected for each GC from the database described in \S \ref{sec:sample}, and the orbital separation is $a = 10 R_{\star} = 4 \rm{R}_{\odot}$. (This is a quite conservative orbital separation. The final circularized radius would typically be smaller than that.) We see that in this case $\lambda$ always presents values below $10^{-2}$, indicating that close stellar encounters with tidally formed compact binaries in the cores of GCs actually make the binaries shrink even more. This seems to be true for all GCs tested here simply because the kinetic temperature of the stars in the cores of GCs is too low (i.e. $\sigma$ is too small).

For comparison, on the right side of Fig. \ref{fig:lambda}, we plot the values of $\lambda$ for the same GCs considering primordial binaries where the two stars have 0.5 M$_{\odot}$ and are separated by a maximum distance equal to the orbital radius of the Earth, $D_{\oplus} =$ 1 AU. Even for these softer binaries, the stellar close encounters will, on average, shrink the systems even more.

We do not observe any significant dependency of the $\gamma$-ray luminosity $L_{\gamma}$ on $\lambda$, meaning that the encounter rate per formed binary may not be important for the evolution of close compact binaries into MSPs. Indeed, the timescales involved in the shrinking of binaries by stellar flybys are very large, sometimes even larger than the Hubble time, as shown by the color bar in Fig. \ref{fig:lambda}, which was computed according to the relation below \citep{quinlan1996dynamical,merritt2001brownian}:

\begin{equation}
    \frac{da}{dt} = \frac{2\pi Gm_{\star}n\xi}{\sigma} a^2 ~~~~~~ \xrightarrow ~~~~~~~ t_{\mathrm{enc}} = \frac{\sigma}{2\pi Gm_{\star}n\xi} \frac{1}{a} \bigg\rvert_{a_0}^{a_f},
    \label{eq:timescale}
\end{equation}
where, for the left side of the plot, $a_f$ is assumed as the tidal radius (i.e. $a_f \approx R_{\star}(2m_{\mathrm{c}}/m_{\star})^{1/3} \approx 1.8R_{\star} \sim 2R_{\star}$), $a_0 = 10 R_{\star}$ is the initial circularized orbital radius, $n$ is the core number density, and $\xi$ is the post-encounter energy parameter, here assumed to be 1. This is a very uncertain dimensionless parameter of order unity that depends on the binary orbital eccentricity, the mass ratio $(m_{\mathrm{c}} + m_{\star})/m_{\star}$, and is averaged over all angular variables describing the binary's orientation and phase \citep[for a detailed discussion on $\xi$, we refer the reader to Eq. 8 in][]{quinlan1996dynamical}. Larger values of $\xi$ imply a larger exchange of energy during the encounters, meaning that the binary would shrink faster. We therefore conclude that stellar close encounters indeed act in favor of shrinking the compact binaries formed in stellar flybys in GCs, however, this is a very inefficient process. These binaries may shrink mainly by other processes, such as angular momentum losses via winds, mass transfer, tides, and/or gravitational waves.

Similarly, for the softer binaries represented on the right side of Fig. \ref{fig:lambda}, we assume $a_0 = D_{\oplus}$ and $a_{f} = a_{\mathrm{GW}}$, which is the orbital radius at which the loss of energy via gravitational waves is already more efficient than the loss by stellar encounters. We compute $a_{\mathrm{GW}}$ by dividing the encounters hardening timescale (Eq. \ref{eq:timescale}) by the gravitational wave orbital decay timescale \citep{peters1964gravitational}, such that:

\begin{equation}
    a_{\mathrm{GW}} = \left( \frac{256G^2m_{\mathrm{c}}(m_{\mathrm{c}}+m_{\star})\sigma}{10\pi \xi (1-e^2)^{7/2}c^5n} \right)^{1/5},
\end{equation}
where $e \equiv 0.1$ is the orbital eccentricity. The shrinkage timescale in both sides of Fig. \ref{fig:lambda} is roughly the same and happens because the smaller the orbital separation $a$, the more difficult it is for the binary to shrink further. Indeed, for 18 of the $\gamma$-ray GCs in our sample, the loss of energy via gravitational waves is already more efficient than stellar encounters at $10R_{\star}$. The smallest value for $a_{\mathrm{GW}}$ is found for NGC 7078 (M15) at $3.5R_{\star}$, while the largest is found for NGC 5139 ($\Omega$ Cen) at $22.0R_{\star}$, such that stellar encounters are much more important for orbital decay in GCs with dynamical properties similar to NGC 7078 than those similar to NGC 5139. We list the dynamical parameters for compact binaries in Table \ref{tab:lambda}.

\begin{table}
    \centering
    \begin{tabular}{l|c|c|c|c}
         Name & $\lambda$ [$10^{-3}$] & $t_{\mathrm{enc}}$ [Gyr] & $a_{\mathrm{GW}}$ [$R_{\star}$] & $\sigma_0$ [km/s]\\
         \hline
        NGC  104  &  2.5  &  216.0  &  12.2  &  11.9 \\
        NGC  1851  &  2.1  &  2.6  &  5.0  &  11.0 \\
        NGC  1904  &  0.7  &  89.5  &  10.2  &  6.5 \\
        NGC  2808  &  3.5  &  165.2  &  11.6  &  14.1 \\
        NGC  362  &  1.3  &  78.2  &  9.9  &  8.6 \\
        NGC  5139  &  5.5  &  4235.3  &  22.1  &  17.7 \\
        NGC  5286  &  1.7  &  63.9  &  9.6  &  9.7 \\
        NGC  5904  &  1.1  &  655.4  &  15.2  &  7.9 \\
        NGC  6093  &  2.1  &  66.1  &  9.6  &  11.0 \\
        NGC  6139  &  2.3  &  21.4  &  7.7  &  11.5 \\
        NGC  6205  &  1.6  &  1828.3  &  18.7  &  9.4 \\
        NGC  6218  &  0.4  &  1514.1  &  18.0  &  4.8 \\
        NGC  6266  &  4.1  &  46.1  &  9.0  &  15.3 \\
        NGC  6304  &  0.5  &  26.9  &  8.0  &  5.5 \\
        NGC  6316  &  1.2  &  170.9  &  11.6  &  8.2 \\
        NGC  6341  &  1.2  &  277.5  &  12.8  &  8.4 \\
        NGC  6342  &  0.3  &  129.5  &  11.0  &  4.4 \\
        NGC  6380  &  1.2  &  680.3  &  15.3  &  8.2 \\
        NGC  6388  &  5.3  &  33.8  &  8.4  &  17.4 \\
        NGC  6397  &  0.5  &  0.9  &  4.1  &  5.3 \\
        NGC  6402  &  1.7  &  2315.0  &  19.6  &  9.9 \\
        NGC  6440  &  3.7  &  39.2  &  8.7  &  14.6 \\
        NGC  6441  &  6.4  &  52.5  &  9.2  &  19.1 \\
        NGC  6517  &  1.7  &  8.8  &  6.4  &  9.9 \\
        NGC  6528  &  0.4  &  115.1  &  10.7  &  4.7 \\
        NGC  6541  &  1.3  &  3.7  &  5.4  &  8.6 \\
        NGC  6637  &  0.7  &  581.1  &  14.9  &  6.1 \\
        NGC  6652  &  0.4  &  56.1  &  9.3  &  4.9 \\
        NGC  6656  &  1.4  &  424.9  &  14.0  &  8.9 \\
        NGC  6681  &  0.9  &  1.4  &  4.5  &  7.1 \\
        NGC  6712  &  0.5  &  1087.6  &  16.8  &  5.1 \\
        NGC  6715  &  6.5  &  54.0  &  9.2  &  19.2 \\
        NGC  6717  &  0.2  &  59.9  &  9.4  &  3.3 \\
        NGC  6723  &  0.5  &  2566.1  &  20.0  &  5.5 \\
        NGC  6752  &  1.0  &  3.8  &  5.4  &  7.6 \\
        NGC  7078  &  3.0  &  0.4  &  3.5  &  13.1 \\
        Terzan 2  &  0.6  &  101.1  &  10.5  &  5.7 \\
        Terzan 5  &  4.2  &  79.3  &  10.0  &  15.5 \\
        Terzan 6  &  1.2  &  59.3  &  9.4  &  8.4 
    \end{tabular}
    \caption{The dynamical parameters for compact binaries in the GCs detected with the \textit{Fermi}-LAT. We see that all GCs present $\lambda \ll 1$, meaning that the secondary stellar encounters will on average make the binaries harder. For 18 GCs, the loss of energy via gravitational waves is already more important than stellar encounters at an orbital distance of 10 $R_{\star}$ (as given by $a_{\mathrm{GW}}$) and the shrinking timescales due to encounters only, $t_{\mathrm{enc}}$, is really large. The last column gives the dispersion velocity in the cores of the GCs.}
    \label{tab:lambda}
\end{table}

\subsection{Looking for extended emission}
\label{sec:results_extended}

The beamed $\gamma$ rays generated in the outer magnetosphere of MSPs in GCs must appear to \textit{Fermi}-LAT as point-like sources since the MSPs are expected to quickly sink to the clusters' cores via dynamical friction (see \S \ref{subsec:extended}). In contrast, relativistic leptons leaking from the magnetospheres of MSPs can propagate within GCs and upscatter ambient thermal photons to the $\gamma$-ray domain \citep{bednarek2007high,zajczyk2013numerical,bednarek2016tev}, similar to what we observe in the surroundings of stars \citep{orlando2008gamma,deMenezes2021superluminous_stars}. This propagation is roughly isotropic, meaning that the resulting $\gamma$-ray emission is independent of the orientation of the MSPs with respect to the observer \citep[see][for a detailed discussion on this topic]{venter2009predictions,bednarek2014misaligned}. This propagation, however, is not well understood and could, in principle, allow the leptons to leave the cores of the clusters before interacting with the soft photons, possibly resulting in extended, and not necessarily spherical, $\gamma$-ray emission \citep{bednarek2014misaligned}. There is a hint for this kind of extended emission (in this case, non-spherical) in the VHE observations of Terzan 5 \citep{abramowski2011terzan5_VHE,ndiyavala2019probing}, but it is still a matter of investigation if the observed $\gamma$-rays are indeed associated to this GC.

In Table \ref{tab:extension} we show the 68\% containment radii (R$_{68}$) for a disk- and a 2D-Gaussian extended emission models applied to the 5 GCs with the largest optical HLRs detected with \textit{Fermi}-LAT, where the HLR here gives us a rough idea of the expected angular size of the GCs if the $\gamma$ rays are generated by the upscattering of thermal photons within the GC. For comparison, we apply the same models to a sample of 100 blazars randomly selected from 4FGL-DR3 (given that blazars are expected to be point-like sources). This sample is divided into five groups, each one containing 20 blazars with similar Galactic latitude ($b$) and $\gamma$-ray flux above 1 GeV ($F_{\gamma}$) to one of the 5 GCs analysed, i.e. $ b_{\mathrm{GC}} ~- ~5^{\circ} < b_{\mathrm{blazar}} < b_{\mathrm{GC}} ~+ ~5^{\circ}$, and $F_{\gamma, \mathrm{GC}}/5 < F_{\gamma, \mathrm{blazar}} < 5\times F_{\gamma, \mathrm{GC}}$. The last four columns of Table \ref{tab:extension} show the median values of TS and median 68\% containment radii for each one of these five groups. We found no strong evidence for extended emission for any of the tested targets. The GCs NGC 6656 and NGC 6752 are those presenting the largest values for R$_{68}$ when compared with the control sample, but even in these cases, extended emission cannot be claimed due to the relatively small values of TS$_{\mathrm{d}}$ and TS$_{\mathrm{G}}$. This is a somewhat expected result in the energy band of \textit{Fermi}-LAT since the bulk of this extended $\gamma$-ray emission is expected to peak at TeV energies \citep{bednarek2007high,venter2009predictions,zajczyk2013numerical}.

\begin{table*}
    \centering
    \begin{tabular}{l|c|c|c|c|c|c|c|c|c}
        Name & TS$_{\mathrm{d}}$ & R$_{68,\mathrm{d}}$ & TS$_{\mathrm{G}}$ & R$_{68,\mathrm{G}}$ & HLR & $\widetilde{\rm{TS}}_{\mathrm{d}}^{20}$ & $\widetilde{\rm{R}}_{68,\mathrm{d}}^{20}$ & $\widetilde{\rm{TS}}_{\mathrm{G}}^{20}$ & $\widetilde{\rm{R}}_{68,\mathrm{G}}^{20}$ \\
        \hline
        NGC 104 & 3.73 & $2.9' \pm 0.6'$ & 3.59 & $2.4' \pm 0.6'$ & $2.8'$ &  1.15  &  $2.3' \pm 0.8'$ &  1.01  &  $2.0' \pm 0.8'$\\
        NGC 5139 & 2.41 & $3.9' \pm 1.2'$ & 2.36 & $3.6' \pm 1.2'$ & $4.8'$ &  1.72  & $ 3.0' \pm  0.9'$ &  1.49  & $ 3.0' \pm  1.0'$\\
        NGC 6397 & 0.82 & $4.8' \pm 3.0'$ & 0.60 & $4.2' \pm 3.6'$ & $3.0'$ &  1.81  & $ 3.9' \pm  0.7'$ &  1.81  & $ 3.7' \pm  0.8'$\\
        NGC 6656 & 1.27 & $8.3' \pm 6.1'$ & 1.39 & $10.0' \pm 7.4'$ & $3.3'$ &  0.45  & $ 1.2' \pm  0.7'$ &  0.51  & $ 1.2' \pm  0.7'$\\
        NGC 6752 & 4.39 & $7.2' \pm 2.4'$ & 4.60 & $7.2' \pm 2.4'$ & $2.4'$ &  0.16  & $ 1.4' \pm  1.1'$ &  0.13  & $ 1.2' \pm  1.2'$
    \end{tabular}
    \caption{Looking for extended emission in the 5 GCs with the largest optical HLRs detected with \textit{Fermi}-LAT. The symbols R$_{68,\mathrm{d}}$ and R$_{68,\mathrm{G}}$ refer to the 68\% containment radii for a disk and a 2D-Gaussian models, respectively, while TS$_{\mathrm{d}}$ and TS$_{\mathrm{G}}$ are their respective values of TS. For each GC, we randomly select a control sample of 20 $\gamma$-ray blazars that have Galactic latitudes and $\gamma$-ray fluxes similar to that of the cluster, and then look for extended emission in all of them (see text for details). The last four columns represent the median values of TS and 68\% confidence radii for these samples. Although the values of R$_{68}$ for all GCs are slightly larger than the median values of the control samples, we have no significant sign of extended $\gamma$-ray emission.}
    \label{tab:extension}
\end{table*}

\subsubsection{The special case of Omega Centauri}

Although we found no significant extended $\gamma$-ray emission from any GC, we noticed a peculiarity when analyzing the RoI of $\Omega$ Centauri (i.e. NGC 5139): there are two patches of extended $\gamma$-ray emission mixed with several UGSs extending northwest and northeast from the cluster. A deeper analysis reveals that these patches and UGSs may actually be present due to residual $\gamma$-ray emission from the extended lobes of the background radio galaxy Centaurus A. Indeed, the observation of this excess $\gamma$-ray emission surrounding Centaurus A has already been discussed in literature \citep{yang2012deep,abdollahi2020_4FGL}, however, it is not considered in the 4FGL catalog due to the lack of an updated morphological study for this source. The current model adopted in 4FGL for the Centaurus A lobes is based on WMAP 22 GHz data \citep{abdo2010_CenA}, which does not account for all of the extended emission seen in $\gamma$ rays. 

\begin{figure*}
    \centering
    \includegraphics[width=\linewidth]{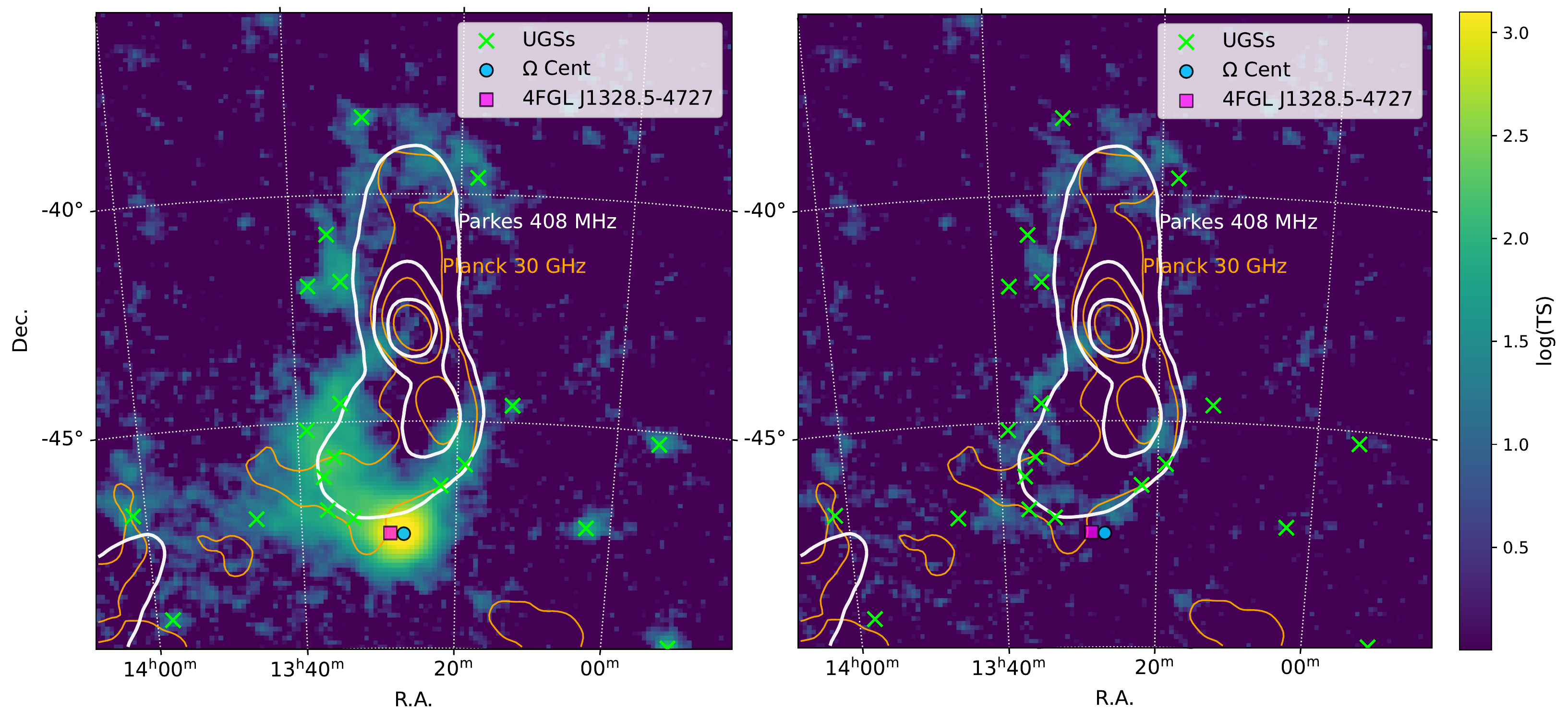}
    \caption{TS maps in the energy range 0.3--300 GeV centered in Centaurus A. Left: RoI model excluding all UGSs, $\Omega$ Centauri/NGC5139 (blue dot at the bottom), and one nearby blazar (magenta square), where we can see the excess $\gamma$-ray emission in the surroundings of Centaurus A. We see that the Parkes (white) and/or Planck (orange) radio contours account for much more of the observed $\gamma$-ray emission, and therefore several of the UGSs in this field (green crosses) are not actually independent point-like $\gamma$-ray sources. Right: Residuals TS map for the same RoI. Even after modeling all of the sources listed in 4FGL-DR3, there is still a blurred excess of TS in the range $10 \lesssim \rm{TS} \lesssim 30$.}
    \label{fig:Cen_A}
\end{figure*}

On the left panel of Fig. \ref{fig:Cen_A}, we show the TS map of an RoI centered on Centaurus A superposed with the radio contours observed with the Parkes telescope at 408 MHz \citep{Haslam1981_Parkes408} and with Planck at 30 GHz\footnote{Observations obtained with Planck (\url{http://www.esa.int/Planck}), an ESA science mission with instruments and contributions directly funded by ESA Member States, NASA, and Canada.}. These contours seem to account for much more of the excess $\gamma$-ray emission than the WMAP 22 GHz template adopted in 4FGL. In this panel, all of the UGSs in the RoI (green crosses), as well as the GC $\Omega$ Centauri (blue dot) and the blazar 4FGL J1328.5-4727 (magenta square), are not included in the model, such that their emission is highlighted in the TS map. On the right panel of Fig. \ref{fig:Cen_A}, we show the same RoI after modeling all sources excluded from the previous plot (i.e. a residuals TS map). Even in this case there is some residual $\gamma$-ray emission related to the lobes, in a TS range of $10 \lesssim \rm{TS} \lesssim 30$. The radio contours adopted here start at 3 times the root mean square of the background and increase in equally logarithmic-spaced intensity intervals.

\subsection{Candidates for radio follow up}

Several GCs detected with the \textit{Fermi}-LAT have no MSP detected in radio (see column ``N$_{\mathrm{P}}$'' in Table \ref{tab:fits_results}), as of January 2023, which can simply happen due to an insufficient radio coverage of these GCs. Indeed, there are also a few clusters for which we have the opposite case: there are pulsars (MSP + young) detected in radio\footnote{\url{http://www.naic.edu/~pfreire/GCpsr.html}} but no significant $\gamma$-ray emission, as shown in Table \ref{tab:GCs_radio_but_no_gamma} and by the magenta upper limits in Fig. \ref{fig:LgxGamma}.

\begin{table}
    \centering
    \begin{tabular}{l|c|c}
        Name & $N_{\mathrm{P}}$ & TS\\
        \hline
        NGC 5024 & 5 & 9 \\
        NGC 5272 & 5 & 4 \\
        NGC 6522 & 6 & 0 \\
        NGC 7089 & 6 & 0 \\
        NGC 7099 & 2 & 0 \\
        Terzan 1* & 7 & 3
    \end{tabular}
    \caption{GCs with pulsars detected in radio ($N_{\mathrm{P}}$) but with no significant $\gamma$-ray emission (TS $<16$). The GC tagged with ``*'' is listed in 4FGL-DR3 as a $\gamma$-ray source, however, in our analysis, it lies outside the 99\% confidence region of the nearest $\gamma$-ray source. We then model and subtract this nearby source from the RoI and are left with an upper limit for Terzan 1 (see the first magenta point on the left in Fig. \ref{fig:LgxGamma}).}
    \label{tab:GCs_radio_but_no_gamma}
\end{table}

Since the $\gamma$-ray and radio emission from MSPs are beamed \citep{abdo2009populationMSPs}, the isotropic $\gamma$-ray luminosity and the number of radio-detected MSPs in a GC are not good representations of the total number of MSPs it contains, but can rather be used to set lower limits on this number. The total number of MSPs in a GC can be more reliably estimated by the following relations:

\begin{equation}
    N_T = \frac{4\pi}{\omega_{\gamma}} \frac{L_{\gamma}}{\langle \dot{E} \rangle \langle\eta_{\gamma}\rangle},
    \label{eq:Ntgamma}
\end{equation}
where $\omega_{\gamma}$ is the typical solid angle covered by the $\gamma$-ray emission from an MSP, $L_{\gamma}$ is the cluster's isotropic $\gamma$-ray luminosity, $\langle \dot{E} \rangle = (1.8\pm 0.7)\times 10^{34}$ erg/s is the average power emitted during the spin-down of MSPs, and $\langle\eta_{\gamma}\rangle = 0.08$ \citep[both values from][]{abdo2010population} is the average efficiency with which the spin-down power is converted into $\gamma$-ray luminosity; and 

\begin{equation}
    N_{\mathrm{T}} = \frac{4\pi}{\omega_{\mathrm{rad}}} N_{\mathrm{rad}},
    \label{eq:Ntradio}
\end{equation}
where $\omega_{\mathrm{rad}}$ is the typical solid angle covered by the radio emission from an MSP and $N_{\mathrm{rad}}$ is the total number of MSPs observed in radio (assuming that the radio observations are deep enough to prevent detection bias). In principle, we don't know what are the appropriate values for $\omega_{\gamma}$ and $\omega_{\mathrm{rad}}$, but we can, however, estimate the fraction $\omega_{\gamma}$/$\omega_{\mathrm{rad}}$ by dividing Eq. \ref{eq:Ntgamma} by Eq. \ref{eq:Ntradio}:

\begin{equation}
    \frac{\omega_{\gamma}}{\omega_{rad}} = \frac{L_{\gamma}}{\langle \dot{E} \rangle \langle\eta_{\gamma}\rangle N_{\mathrm{rad}}}.
    \label{eq:fraction_angle}
\end{equation}
This fraction tells us how large the $\gamma$-ray emitting region is if compared to the radio-emitting region. If we use the parameters of NGC 104 in Eq. \ref{eq:fraction_angle}, we get 

\begin{equation}
    \left.\frac{\omega_{\gamma}}{\omega_{rad}}\right|_{\mathrm{NGC104}} = 1.4\pm 0.9.
\end{equation}
We focus on NGC 104 because it has one of the largest populations of radio-detected MSPs (see Table \ref{tab:fits_results}), has the highest significance among the GCs observed by \textit{Fermi}-LAT, is one of the closest GCs to the Solar System, and is quite far from the Galactic disk \citep[for more details on the observations of MSPs in NGC 104, we refer the reader to][]{Camilo2000_47_Tuc,Freire2001_47_Tuc}. Finding a precise value for $\omega_{\gamma}$/$\omega_{\mathrm{rad}}$ is a hard task since this value can substantially change if we use the parameters of different GCs, mostly due to the uneven radio observations of these clusters. In any case, we take NGC 104 as our reference, implying that the solid angle covered by the $\gamma$-ray emission in an MSP is $1.4\pm 0.9$ times larger than the solid angle covered by the radio beam (and of course, it has to be $< 4\pi$).

We use this value to estimate the number of radio-detectable MSPs that should be observed in the 5 brightest GCs detected by LAT that have no radio pulsars yet observed, as shown in Table \ref{tab:GC_radio_survey}. These clusters are excellent targets for radio campaigns, especially NGC 6388 and NGC 6316.

\begin{table}
    \centering
    \begin{tabular}{c|l}
       Name  & N$_{\mathrm{expec}}$ \\
       \hline
       NGC 2808  &  $2\sim6$ \\
       NGC 6093  &  $2\sim7$ \\
       NGC 6316  &  $5\sim22$  \\
       NGC 6388  &  $9\sim40$  \\
       NGC 6541  &  $1\sim4$
    \end{tabular}
    \caption{The expected number of detectable radio MSPs in the brightest GCs detected with the \textit{Fermi}-LAT that have no pulsars detected in radio to date. All these GCs are excellent targets for a radio follow-up.}
    \label{tab:GC_radio_survey}
\end{table}

\section{Discussion}
\label{sec:discussion}

Stellar close encounters represent the major factor in the formation of close compact binaries in a GC. The recipe for a large population of MSPs requires a GC with high stellar density and low dispersion velocity, as well as a relatively high number of neutron stars retained in the cluster after the supernova explosions. Another factor that could exert some impact on the size of the population of MSPs is the thermodynamic state of the GC's core: if the dynamical interactions are dominated by stellar-mass black holes, the core should be bloated (i.e. non-core-collapsed) and inhibit the mass-segregation of neutron stars to the GC's core, suppressing the formation of MSPs \citep{claire2019rate}. Only after most of the stellar-mass black holes are ejected from the cluster (i.e. a core-collapsed cluster) the neutron stars can sink to the center and interact with other stars to form compact binary systems \citep{claire2019rate,kremer2019initial}. In this work, however, we do not observe any favoritism for a higher number of X-ray sources or higher $\gamma$-ray luminosity in the core-collapsed GCs, although we observe that they present the shortest binary shrinking timescales due to secondary stellar encounters (e.g. see the five darkest points in Fig. \ref{fig:X-ray-EncRate}). 

In this work, the X-ray observations proved to be much more useful to probe the formation of compact binaries than $\gamma$ rays in the GeV band, and this may be due to the fact that we can reliably estimate the total number of X-ray sources in a GC, while in $\gamma$ rays we have access only to the luminosity, which can significantly fluctuate independently of the dynamical parameters of the GCs, making the $L_{\gamma} \times \Gamma$ correlation more dispersed. Another possibility for this higher dispersion may be that in some GCs the compact binaries do not easily evolve into MSPs. This could happen, for instance, if for any reason (possibly related to the cluster's initial mass function) a GC has an overabundance of stellar-mass black holes or white dwarfs if compared to the number of neutron stars, such that most of the compact binaries do not have a neutron star that can be spun up until becoming an MSP. This would not affect the total number of X-ray sources, but could affect the $\gamma$-ray luminosity.

\subsection{The impossibility of compact binary ionization in GCs}
\label{sec:impossible_ionization}

Another interesting point to highlight about the left side of Fig. \ref{fig:lambda} is that if we arbitrarily increase the dispersion velocity ($\sigma$) of the stars in the core of the cluster until it reaches the central escape velocity, then the values of the normalized encounter rate per formed binary of the 39 GCs will range between $0.003 < \lambda < 0.104$, meaning that stellar encounters will never be able to ionize compact binary systems in the cores of these GCs. This happens simply because the fastest stars, i.e., those with higher chances of ionizing/exciting compact binaries, will tend to escape from the clusters' gravitational potential well.

\section{Conclusions}
\label{sec:conclusions}

In this work, we investigated how the dynamical properties of GCs affect the formation and evolution of compact binary systems. We used \textit{Chandra} X-ray and \textit{Fermi}-LAT $\gamma$-ray data to measure the effect of close stellar encounters in the total number of compact binaries and eventually MSPs that we can find in a GC. We identified 39 GCs in the energy range 0.3 -- 300 GeV, seven of which are not listed as GCs in 4FGL-DR3, namely NCC 6342, NGC 6380, NGC 6517, NGC 6528, NGC 6637, NGC 6723, and Terzan 6. Our main results are summarized below.

\begin{enumerate}
    \item We measured the $\gamma$-ray flux and spectral parameters for all 39 GCs, as listed in Table \ref{tab:fits_results} and found that all of them present pulsar-like characteristics. This is even more clear in Fig. \ref{fig:colorcolor}, where we see the distribution of GCs in the $E_{\mathrm{peak}} \times \alpha$ space. The distribution of different classes of astrophysical sources in this figure can also be used in future works as, for instance, the association of $\gamma$-ray sources with their low-energy counterparts \citep[especially on a machine learning framework, similar to the method described in][]{deMenezes2020physical_assoc}.
    \item We found that the total number of X-ray sources is tightly correlated with the stellar encounter rate in GCs with $\Gamma > 100$, which implies that stellar encounters are a major formation channel for compact binaries in GCs. The correlation is tighter in X-rays, where we can reliably estimate the total number of sources in each GC, while in $\gamma$ rays we have access only to the clusters' luminosity, which can significantly fluctuate and make the correlation more dispersed. In GCs with $\Gamma < 100$, we observe an excess in the number of X-ray sources, indicating that another formation channel for compact binaries may be evoked, or even that these clusters favor the formation of non-binary X-ray sources.     
    \item Regardless of the formation channel, secondary stellar encounters in GCs will, on average, make the compact binaries shrink even more. This happens because the kinetic temperature of stars in Galactic GCs is not high enough to ionize these systems, as investigated in the framework of the Heggie-Hills law. This shrinking process, however, is not efficient, as the involved timescales are too long. Furthermore, we argue that the ionization of compact binary systems by stellar encounters in the cores of the 39 GCs studied here may actually be impossible (see \S \ref{sec:impossible_ionization}).
    \item We found no extended $\gamma$-ray emission from the GCs, indicating that the pulsars are concentrated in their cores, probably due to dynamical friction. Extended $\gamma$-ray emission is also possible in the $\gamma$-ray domain, depending on the propagation of leptons leaking from the MSPs magnetospheres within the GCs.
    \item There is an excess of $\gamma$-ray emission coming from the lobes of Cen A extending beyond the WMAP-based model adopted in 4FGL. We found that Planck 30 GHz and Parkes 408 MHz data have a better spatial agreement with the \textit{Fermi}-LAT data, although they still seem insufficient to explain all of the observed extended $\gamma$-ray emission.
    \item We did a preliminary investigation of the rate $\omega_{\gamma}/\omega_{rad}$, that gives us an idea of the opening angle of the radio and $\gamma$-ray emission coming from the magnetospheres of MSPs. A precise estimate of this ratio can help us in estimating the total number of MSPs in GCs (as stated in Eqs. \ref{eq:Ntgamma} and \ref{eq:Ntradio}).
    \item We selected 5 GCs as excellent targets for radio follow-ups (Table \ref{tab:GC_radio_survey}) since they present a substantially high $\gamma$-ray emission but no MSP detected in radio so far.
\end{enumerate}

Besides these results, we leave an open question on what kind of phenomenon, other than close stellar encounters, could induce the formation of X-ray sources in GCs, since we observed an unexpected increase in the number of X-ray sources for GCs with $\Gamma < 100$.

\section*{Acknowledgements}

We would like to thank Teddy Cheung, who told us that the excess of $\gamma$-ray emission on the surroundings of $\Omega$ Centauri could actually be related to the lobes of the radio galaxy Centaurus A, and Giacomo Principe for the very helpful comments and suggestions that substantially improved the robustness of our results. We furthermore thank Raffaele D'Abrusco, David Smith, Guillem Marti-Devesa, Matthew Kerr, and Melissa Pesce-Rollins for the comments along the development of this work, as well as the anonymous referee for the very interesting comments and suggestions. R.M. acknowledges support from the Universit\`a degli Studi di Torino and Istituto Nazionale di Fisica Nucleare (INFN), Sezione di Torino, under the assegno di ricerca Sviluppo di sensori basati su SiPM per ricerca di sorgenti di fotoni di $E>100$ GeV (finanziamento MIUR Dipartimenti di Eccellenza 2018-2022 – per il Dipartimento di Fisica), DFI.2021.24. We acknowledge the use of NASA's SkyView facility (\url{http://skyview.gsfc.nasa.gov}) located at NASA Goddard Space Flight Center. This research has made use of data obtained from the Chandra Source Catalog, provided by the Chandra X-ray Center (CXC) as part of the Chandra Data Archive.

The \textit{Fermi}-LAT Collaboration acknowledges generous ongoing support
from a number of agencies and institutes that have supported both the
development and the operation of the LAT as well as scientific data analysis.
These include the National Aeronautics and Space Administration and the
Department of Energy in the United States, the Commissariat \`a l'Energie Atomique
and the Centre National de la Recherche Scientifique / Institut National de Physique
Nucl\'eaire et de Physique des Particules in France, the Agenzia Spaziale Italiana
and the Istituto Nazionale di Fisica Nucleare in Italy, the Ministry of Education,
Culture, Sports, Science and Technology (MEXT), High Energy Accelerator Research
Organization (KEK) and Japan Aerospace Exploration Agency (JAXA) in Japan, and
the K.~A.~Wallenberg Foundation, the Swedish Research Council and the
Swedish National Space Board in Sweden.

Additional support for science analysis during the operations phase is gratefully
acknowledged from the Istituto Nazionale di Astrofisica in Italy and the Centre
National d'\'Etudes Spatiales in France. This work performed in part under DOE
Contract DE-AC02-76SF00515.

\section*{Data Availability}

The data used in this work can be found online in the \textit{Fermi}-LAT data server\footnote{\url{https://fermi.gsfc.nasa.gov/cgi-bin/ssc/LAT/LATDataQuery.cgi}}, in the globular cluster database\footnote{\url{https://people.smp.uq.edu.au/HolgerBaumgardt/globular/}}, in the online list of pulsars in GCs\footnote{\url{http://www.naic.edu/~pfreire/GCpsr.html}}, and in the Chandra CSC 2.0. online database \url{https://cxc.cfa.harvard.edu/csc/}. Planck and Parkes's observations were collected with SkyView (\url{http://skyview.gsfc.nasa.gov}). The data analysis was performed with \texttt{Fermitools}, \texttt{fermipy}, and \texttt{easyFermi}, all of which are free to download and use (see \S \ref{sec:observations}).



\bibliographystyle{mnras}
\bibliography{main} 





\appendix

\section{Upper limits}
\label{appendix}

In Table \ref{tab:ULs}, we show the integrated energy flux (0.3 -- 300 GeV) upper limits for those GCs in our sample presenting TS $<16$. The adopted spectral model is the PLEC described in Eq. \ref{eq:PLEC}, where $E_{\mathrm{c}} = 3$ GeV and $\alpha = 2$.

\begin{table}
    \centering
    \begin{tabular}{l|c|c|c|c|c|c|c}
         Name & $\Gamma$ & N$_{\mathrm{P}}$ & TS & eFlux UL 95\% \\
  &   &   &   & $10^{-13}$ erg cm$^{-2}$ s$^{-1}$ \\
\hline
am1  & $0.00419^{+ 0.00464 }_{- 0.0017 }$ &  0  &  0.0  &  1.99  \\
arp2  & $0.00518^{+ 0.00285 }_{- 0.00188 }$ &  0  &  5.95  &  10.12  \\
hp  & $0.662^{+ 0.407 }_{- 0.303 }$ &  0  &  1.05  &  22.22  \\
ic1276  & $7.97^{+ 7.99 }_{- 3.71 }$ &  0  &  0.0  &  7.88  \\
ic4499  & $0.797^{+ 0.412 }_{- 0.272 }$ &  0  &  7.61  &  9.52  \\
NGC  1261  & $15.4^{+ 10.6 }_{- 4.3 }$ &  0  &  0.0  &  1.05  \\
NGC  2298  & $4.31^{+ 1.48 }_{- 1.21 }$ &  0  &  0.0  &  1.97  \\
NGC  2419  & $2.8^{+ 0.754 }_{- 0.532 }$ &  0  &  0.0  &  2.59  \\
NGC  288  & $0.766^{+ 0.284 }_{- 0.205 }$ &  0  &  0.0  &  1.02  \\
NGC  3201  & $7.17^{+ 3.56 }_{- 2.27 }$ &  0  &  4.12  &  8.45  \\
NGC  4147  & $16.6^{+ 12.5 }_{- 6.36 }$ &  0  &  0.0  &  2.74  \\
NGC  4372  & $0.233^{+ 0.365 }_{- 0.124 }$ &  0  &  8.3  &  12.68  \\
NGC  4590  & $5.82^{+ 2.69 }_{- 1.7 }$ &  0  &  0.0  &  1.97  \\
NGC  5024  & $35.4^{+ 12.4 }_{- 9.6 }$ &  5  &  1.1  &  4.04  \\
NGC  5053  & $0.105^{+ 0.061 }_{- 0.0362 }$ &  0  &  6.68  &  6.73  \\
NGC  5272  & $194.0^{+ 33.1 }_{- 18.0 }$ &  6  &  4.46  &  4.94  \\
NGC  5466  & $0.239^{+ 0.0666 }_{- 0.0475 }$ &  0  &  0.31  &  2.79  \\
NGC  5634  & $20.2^{+ 14.2 }_{- 7.5 }$ &  0  &  0.0  &  1.58  \\
NGC  5694  & $191.0^{+ 52.2 }_{- 34.4 }$ &  0  &  6.5  &  8.58  \\
NGC  5824  & $984.0^{+ 171.0 }_{- 155.0 }$ &  0  &  0.0  &  2.68  \\
NGC  5897  & $0.851^{+ 0.357 }_{- 0.189 }$ &  0  &  0.0  &  3.83  \\
NGC  5927  & $68.2^{+ 12.7 }_{- 10.3 }$ &  0  &  1.23  &  10.44  \\
NGC  5946  & $134.0^{+ 33.6 }_{- 44.6 }$ &  0  &  0.0  &  5.07  \\
NGC  5986  & $61.9^{+ 15.9 }_{- 10.4 }$ &  1  &  0.0  &  1.04  \\
NGC  6101  & $0.974^{+ 0.567 }_{- 0.287 }$ &  0  &  0.0  &  2.55  \\
NGC  6121  & $26.9^{+ 11.6 }_{- 9.56 }$ &  1  &  0.54  &  8.95  \\
NGC  6144  & $3.14^{+ 1.07 }_{- 0.85 }$ &  0  &  0.0  &  2.09  \\
NGC  6171  & $6.77^{+ 2.34 }_{- 1.72 }$ &  0  &  0.0  &  2.14  \\
NGC  6229  & $47.6^{+ 31.0 }_{- 9.36 }$ &  0  &  0.0  &  1.67  \\
NGC  6235  & $5.75^{+ 2.72 }_{- 1.59 }$ &  0  &  4.66  &  13.56  \\
NGC  6254  & $31.4^{+ 4.34 }_{- 4.08 }$ &  2  &  0.42  &  7.57  \\
NGC  6256  & $169.0^{+ 119.0 }_{- 60.4 }$ &  0  &  0.38  &  12.13  \\
NGC  6273  & $200.0^{+ 66.6 }_{- 38.6 }$ &  0  &  0.0  &  3.49  \\
NGC  6284  & $666.0^{+ 122.0 }_{- 105.0 }$ &  0  &  0.0  &  1.55  \\
NGC  6287  & $36.3^{+ 7.7 }_{- 7.74 }$ &  0  &  1.7  &  11.53  \\
NGC  6293  & $847.0^{+ 377.0 }_{- 239.0 }$ &  0  &  2.97  &  13.81  \\
NGC  6325  & $118.0^{+ 44.7 }_{- 45.6 }$ &  0  &  1.55  &  13.25  \\
NGC  6333  & $131.0^{+ 59.1 }_{- 41.8 }$ &  0  &  0.0  &  2.12  \\
NGC  6352  & $6.74^{+ 1.71 }_{- 1.3 }$ &  0  &  3.34  &  12.33  \\
NGC  6355  & $99.2^{+ 41.1 }_{- 25.7 }$ &  0  &  0.0  &  7.81  \\
NGC  6356  & $88.1^{+ 20.2 }_{- 13.7 }$ &  0  &  0.0  &  5.75  \\
NGC  6362  & $4.56^{+ 1.51 }_{- 1.03 }$ &  0  &  0.0  &  2.09     
    \end{tabular}
    \caption{Integrated energy flux 95\% upper limits in the energy range 0.3 -- 300 GeV for the GCs with TS $<16$. The three GCs tagged with ``*'' are detected by \textit{Fermi}-LAT, although their emission are attributed to individual MSPs, as mentioned in the main text. We also provide the stellar encounter rate $\Gamma$, the number of detected pulsars in radio $N_{\mathrm{P}}$, and the TS for each source.}
    \label{tab:ULs}
\end{table}

\begin{table}
    \ContinuedFloat 
    \centering
    \begin{tabular}{l|c|c|c|c|c|c|c}
         Name & $\Gamma$ & N$_{\mathrm{P}}$ & TS & eFlux UL 95\% \\
  &   &   &   & $10^{-13}$ erg cm$^{-2}$ s$^{-1}$ \\
\hline
NGC  6366  & $5.14^{+ 2.75 }_{- 1.76 }$ &  0  &  0.0  &  6.5  \\
NGC  6401  & $44.0^{+ 11.0 }_{- 10.7 }$ &  0  &  7.77  &  26.59  \\
NGC  6426  & $1.58^{+ 1.09 }_{- 0.518 }$ &  0  &  0.3  &  7.19  \\
NGC  6453  & $371.0^{+ 128.0 }_{- 88.7 }$ &  0  &  8.27  &  22.79  \\
NGC  6496  & $0.657^{+ 0.616 }_{- 0.289 }$ &  0  &  0.0  &  4.29  \\
NGC  6522  & $363.0^{+ 113.0 }_{- 98.5 }$ &  5  &  6.83  &  22.28  \\
NGC  6535  & $0.388^{+ 0.389 }_{- 0.192 }$ &  0  &  8.03  &  17.95  \\
NGC  6539  & $42.1^{+ 28.6 }_{- 15.3 }$ &  1  &  0.0  &  26.66  \\
NGC  6544  & $111.0^{+ 67.8 }_{- 36.5 }$ &  2  &  7.0  &  60.7  \\
NGC  6553  & $69.0^{+ 26.8 }_{- 18.8 }$ &  0  &  4.46  &  21.57  \\
NGC  6558  & $105.0^{+ 26.2 }_{- 19.3 }$ &  0  &  0.0  &  3.73  \\
NGC  6569  & $53.6^{+ 30.2 }_{- 20.8 }$ &  0  &  0.02  &  7.35  \\
NGC  6584  & $11.8^{+ 5.39 }_{- 3.36 }$ &  0  &  0.0  &  2.19  \\
NGC  6624*  & $1150.0^{+ 113.0 }_{- 178.0 }$ &  12  &  0.0  &  27.34 \\
NGC  6626*  & $648.0^{+ 83.8 }_{- 91.1 }$ &  14  &  7.62  &  25.35  \\
NGC  6637  & $89.9^{+ 36.0 }_{- 18.1 }$ &  0  &  11.0  &  8.29 \\
NGC  6638  & $137.0^{+ 38.6 }_{- 27.1 }$ &  0  &  2.81  &  12.88  \\
NGC  6642  & $97.8^{+ 31.3 }_{- 24.5 }$ &  0  &  0.18  &  7.99  \\
NGC  6652*  & $700.0^{+ 292.0 }_{- 189.0 }$ &  2  &  0.0  &  43.28  \\
NGC  6760  & $56.9^{+ 26.6 }_{- 19.4 }$ &  2  &  0.07  &  9.11  \\
NGC  6779  & $27.7^{+ 12.2 }_{- 9.16 }$ &  0  &  2.03  &  7.46  \\
NGC  6809  & $3.23^{+ 1.38 }_{- 1.0 }$ &  0  &  0.28  &  5.79  \\
NGC  6864  & $307.0^{+ 93.5 }_{- 89.3 }$ &  0  &  2.77  &  7.84  \\
NGC  6934  & $29.9^{+ 12.0 }_{- 8.22 }$ &  0  &  3.2  &  7.61  \\
NGC  6981  & $4.69^{+ 2.52 }_{- 1.76 }$ &  0  &  0.0  &  1.59  \\
NGC  7006  & $9.4^{+ 4.92 }_{- 3.25 }$ &  0  &  3.47  &  7.24  \\
NGC  7089  & $518.0^{+ 77.6 }_{- 71.4 }$ &  6  &  0.77  &  4.43  \\
NGC  7099  & $324.0^{+ 124.0 }_{- 81.2 }$ &  2  &  7.8  &  8.16  \\
NGC  7492  & $0.192^{+ 0.243 }_{- 0.0765 }$ &  0  &  7.69  &  7.11  \\
Pal  1  & $0.895^{+ 0.6 }_{- 0.241 }$ &  0  &  0.0  &  1.39  \\
Pal  10  & $59.0^{+ 42.8 }_{- 35.5 }$ &  0  &  0.0  &  2.95  \\
Pal  11  & $20.8^{+ 11.2 }_{- 7.11 }$ &  0  &  0.0  &  2.24  \\
Pal  12  & $0.397^{+ 0.42 }_{- 0.216 }$ &  0  &  4.35  &  6.13  \\
Pal  13  & $0.00109^{+ 0.00168 }_{- 0.00062 }$ &  0  &  0.0  &  2.84  \\
Pal  14  & $0.00186^{+ 0.000872 }_{- 0.000459 }$ &  0  &  0.27  &  4.18  \\
Pal  2  & $929.0^{+ 836.0 }_{- 555.0 }$ &  0  &  0.26  &  7.96  \\
Pal  3  & $0.0409^{+ 0.0192 }_{- 0.00984 }$ &  0  &  1.23  &  4.5  \\
Pal  4  & $0.0189^{+ 0.0109 }_{- 0.00366 }$ &  0  &  0.0  &  1.79  \\
Pal  5  & $0.00212^{+ 0.000861 }_{- 0.000627 }$ &  0  &  0.0  &  3.69  \\
Pal  6  & $15.5^{+ 13.2 }_{- 7.75 }$ &  0  &  7.44  &  35.96  \\
Pal  8  & $4.22^{+ 2.66 }_{- 1.21 }$ &  0  &  2.6  &  13.32  \\
Terzan  1  & $0.292^{+ 0.274 }_{- 0.17 }$ &  7  &  0.0  &  82.06  \\
Terzan  7  & $1.59^{+ 1.09 }_{- 0.634 }$ &  0  &  0.0  &  2.11  \\
Terzan  9  & $1.71^{+ 1.67 }_{- 0.959 }$ &  0  &  0.0  &  9.36  \\
Terzan  2  & $4.29^{+ 3.72 }_{- 1.73 }$ &  0  &  3.95  &  21.15  
    \end{tabular}
    \caption{Continued.}
\end{table}



\bsp	
\label{lastpage}
\end{document}